\documentclass[twocolumn,pdff,prb,amsmath,amssymb,showpacs,floatfix]{revtex4}
\usepackage{graphicx}
\usepackage{color}

\begin{document}
\title{Chiral states in bilayer graphene: magnetic field dependence and gap opening}

\author{M. Zarenia$^1$, J. M. Pereira Jr.$^2$, G. A. Farias$^2$, and F. M. Peeters$^{1,2}$}
\address{$^1$Department of Physics, University of Antwerp,
Groenenborgerlaan 171, B-2020 Antwerpen, Belgium.\\
$^2$Departamento de F\'{\i}sica, Universidade Federal do Cear\'a,
Fortaleza, Cear\'a, 60455-760, Brazil.}

\begin{abstract}
At the interface of electrostatic potential kink profiles one
dimensional chiral states are found in bilayer graphene (BLG). Such
structures can be created by applying an asymmetric potential to the
upper and the lower layer of BLG. We found that: \emph{i}) due to
the strong confinement by the single kink profile the
uni-directional states are only weakly affected by a magnetic field,
\emph{ii}) increasing the smoothness of the kink potential results
in additional bound states which are topologically different from
those chiral states, and \emph{iii}) in the presence of a
kink-antikink potential the overlap between the oppositely moving
chiral states results in the appearance of crossing and
anti-crossing points in the energy spectrum. This leads to  the
opening of tunable minigaps in the spectrum of the uni-directional
topological states.
\end{abstract}

 \maketitle

\section{Introduction}
Carbon-based electronic structures have been the focus of intense
research since the discovery of fullerenes and carbon nanotubes
\cite{Millie}. More recently, the production of atomic layers of
hexagonal carbon (called graphene) has renewed that interest, with
the observation of striking mechanical and electronic properties, as
well as ultrarelativistic-like phenomena in this condensed matter
system\cite{Review}. In that context, bilayer graphene (BLG), which
is a system with two Van der Waals coupled sheets of graphene, has
been shown to have features that make it a possible substitute of
silicon in microelectronic devices. The carrier dispersion of
pristine BLG is gapless and approximately parabolic at two points in
the Brillouin zone ($K$ and $K'$). However, it was found that the
application of perpendicular electric fields produced by external
gates deposited on the BLG surface can induce a gap in the spectrum,
by creating a charge imbalance between the two graphene
layers\cite{Mccann,Castro}. The tailoring of the gap by an external
field may be particularly useful for the development of devices
\cite{Milton1,zarenia}.
\begin{figure}
\centering
\includegraphics[width=7cm]{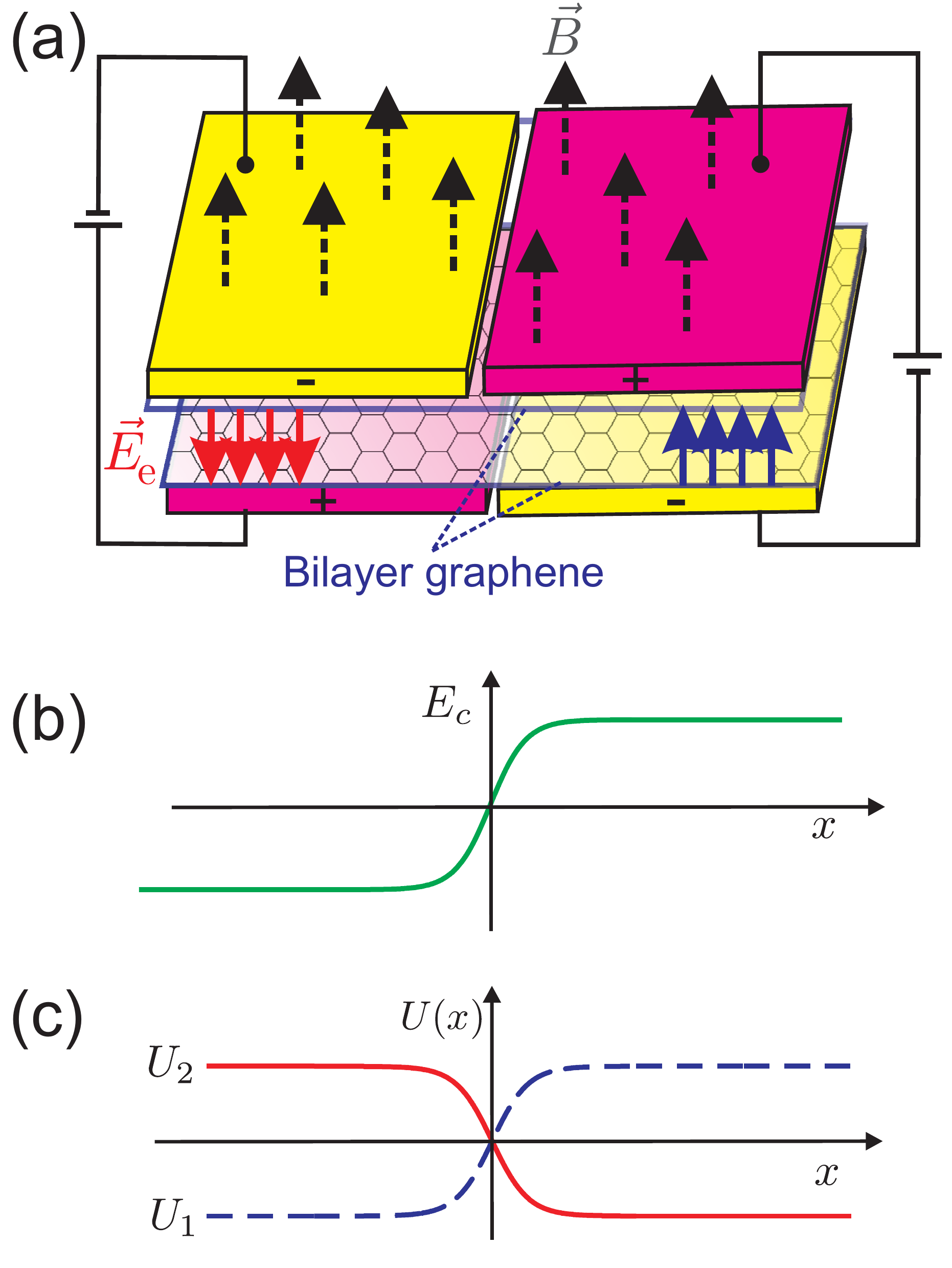}
 \caption{(Color online)
(a) Schematic illustration of the non-uniformly gated bilayer
graphene device for the creation of a kink potential. Applied gated
voltage to the upper and lower layers with opposite sign induces an
electric field $\boldsymbol{E_{e}}$, with preferential direction. An
external magnetic field $\boldsymbol{B}=B~\hat{z}$, is applied
perpendicular to the bilayer graphene sheets. (b) Electric field
between the two graphene layers. (c) Potential on layer 1 ($U_{1}$)
and layer 2 ($U_{2}$).} \label{fig0}
\end{figure}
It was recently recognized that a tunable energy gap in BLG can allow the observation
of new confined electronic states, which could be obtained by applying a spatially
varying potential profile to create a position-dependent gap analogous to semiconductor heterojunctions.

An alternative way to create one dimensional localized states in BLG
has recently been suggested by Martin {\it et al.} \cite{Morpurgo}
and relies on the creation of a potential "kink" by an asymmetric
potential profile (see Fig. \ref{fig0}). Such kink potential can
also be realized in p-n junctions. They showed that localized chiral
states arise at the location of the kink, with energies inside the
energy gap. These states correspond to unidirectional motion of
electrons which are analogous to the edge states in a quantum Hall
system. From a practical standpoint, the kinks may be envisaged as
configurable metallic nanowires embedded in a semiconductor medium.
Moreover, the carrier states in this system are expected to be
robust with regards to scattering and may display Luttinger liquid
behavior \cite{Killi}.

\indent An additional tool for the manipulation of charged states in
BLG is the use of magnetic fields. The application of an external
magnetic field perpendicular to the BLG sheet causes the appearance
of Landau levels which can be significantly modified by the induced
gap, leading to effects such as the lifting of valley degeneracy
caused by the breaking of the inversion symmetry by the
electrostatic bias \cite{Falko,Milton2}. Recently the transport
properties of p-n-p junctions in bilayer graphene were
experimentally investigated in the presence of a perpendicular
magnetic field\cite{Jing}.

In the present paper we generalize previous work on topological
confinement in bilayer graphene on three levels: \emph{i}) we
investigate the effect of smoothing the kink potential on the
topological states, \emph{ii}) the effect of a perpendicular
magnetic field is studied, and \emph{iii}) we investigate a new
system that consists of a coupled kink-antikink structure. We
demonstrate that the latter opens a gap in the 1D electron states.
The paper is organized as follows. In Sec. II we present the
theoretical formalism. The results for a single kink potential
profile are discussed in Sec. III(A,B). In Sec. IV(A) and Sec. IV(B)
we show the results for the kink-antink potential, respectively, for
zero and non-zero magnetic fields. Finally we conclude the remarks
of the paper in Sec. V.

\section{Model}
We employ a two-band continuum model to describe the BLG sheet. In this model, the system is
described by four sublattices, two in the upper ($A$, $B$) and two in the lower ($A'$ and $B'$)
layer\cite{Milton1}. The interlayer coupling is given
by the hopping parameter $t \approx 400~meV$ between sites $A$ and $B'$.
The Hamiltonian around the $K$ valley of the first Brillouin zone
can be written as
\begin{equation}\label{H}
H=-\frac{1}{t}\left[
  \begin{array}{cc}
  0 &(\pi^{\dag})^{2}\\
  (\pi)^{2}& 0\\
  \end{array}
\right]+\left[
  \begin{array}{cc}
  U(x) &0\\
  0& -U(x)\\
  \end{array}
\right]
\end{equation}
where $\pi = v_F(p_x + i p_y)$, $p_{x,y} = - i \hbar
\partial_{x,y}+eA_{x,y}$ is the momentum operator in the presence of
an external magnetic field with $A_{x,y}$ being the components of
the vector potential $\mathbf{A}$, $v_F = 10^6$ m/s is the Fermi
velocity, $U(x)$ and $-U(x)$ is the electrostatic potential
respectively applied to the upper and lower layers. The eigenstates
of the Hamiltonian Eq. (1) are two-component spinors $\Psi(x,y) =
[\psi_a(x,y) ,\psi_b(x,y)]^T$, where $\psi_{a,b}$ are the envelope
functions associated with the probability amplitudes at sublattices
$A$ and $B'$ at the respective layers of the BLG sheet. Since
$[H,p_{y}]=0$ the momentum along the $y$-direction is a conserved
quantity and therefore we can write:
\begin{equation}\label{Psi}
\displaystyle{\psi(x,y)=e^{ik_{y}y}[\varphi_{a}(x),\varphi_{b}(x)]^{T}}
\end{equation}
where, $k_{y}$ is the wave vector along the $y$ direction.

In order to apply a perpendicular magnetic field to the bilayer
sheet we employ the Landau gauge for the vector potential
$\mathbf{A}=(0,Bx,0)$. The Hamiltonian (\ref{H}) acts on the wave
function of Eq. (\ref{Psi}) which leads to the following coupled
second-order differential equations,
\begin{subequations}\label{eqS}
\begin{eqnarray}
&&\big[\frac{\partial}{\partial x'}+(k'_{y}+\beta x')\big]^{2}\varphi_{b}=[\epsilon-u(x')]\varphi_{a}, \\
&&\big[\frac{\partial}{\partial x'}-(k'_{y}+\beta x')\big]^{2}\varphi_{a}=[\epsilon+u(x')]\varphi_{b}.
\end{eqnarray}
\end{subequations}
where, in the above equations we used dimensionless units $x'=x/l$,
$k'_{y}=k_{y}l$, $\epsilon=E/t$, and $u(x')=U(x)/t$ where $l=\hbar
v_{F}/t=1.6455~nm$, $\beta=(eB/\hbar)l^{2}~(=0.0041$ for $B=1~T)$.
The step-like kink (see Fig. 1(c)) is modeled by
\begin{equation}
\displaystyle{u(x')=u_{b}\tanh(x'/\delta)},
\end{equation}
where $u_{b}$ is the maximum value of the gate voltage, in
dimensionless unit, in each BLG layer. Here, $\delta$ denotes the
width of the region in which the potential switches its sign in each
layer. This parameter is determined by the distance between the
gates used to create the gap. Next, we numerically solve Eqs. (3) to
obtain the dependence of the energy levels on the magnetic field and
potential parameters. For the case of a sharp kink potential
$\delta\rightarrow0$ and in the absence of a magnetic field , i.e.
$B=0$, Eqs. (\ref{eqS}) reduces to
\begin{subequations}\label{eqSB0}
\begin{eqnarray}
&&\big[\frac{\partial}{\partial x'}+k'_{y}\big]^{2}\varphi_{b}=[\epsilon-u(x')]\varphi_{a}, \\
&&\big[\frac{\partial}{\partial x'}-k'_{y}\big]^{2}\varphi_{a}=[\epsilon+u(x')]\varphi_{b}.
\end{eqnarray}
\end{subequations}
where $u(x')=u_{b}[\Theta(x')-\Theta(-x')]$. We simply decouple Eqs. (\ref{eqSB0}) and obtain
\begin{equation}
[\frac{\partial^{2}}{\partial x^{'2}}+\lambda_{\pm}^{2}]\varphi_{a}=0
\end{equation}
where,
$\lambda_{\pm}=\big[-k_{y}^{'2}\pm\sqrt{\epsilon^{2}-u_{b}^{2}}\big]^{1/2}$
which can be a complex quantity. The solution for $x'<0$
($\psi^{<}$) and $x'\geq0$ ($\psi^{>}$) are given by
\begin{subequations}\label{eqi}
\begin{eqnarray}
&&\psi(x')^{<}_{\pm}=\left(
             \begin{array}{c}
                e^{i\lambda_{\pm}x'} \\
              f_{\pm}e^{i\lambda_{\pm}x'} \\
             \end{array}
           \right)
,\\
&&\psi(x')^{>}_{\pm}=\left(
             \begin{array}{c}
               e^{-i\lambda_{\pm}x'} \\
               g_{\pm}e^{- i\lambda_{\pm}x'} \\
             \end{array}
           \right)
\end{eqnarray}
\end{subequations}
where, $f_{\pm}=( i\lambda_{\pm}-k'_{y})^{2}/(\epsilon-u_{b})$, and
$g_{\pm}=(i\lambda_{\pm}+k'_{y})^{2}/(\epsilon+u_{b})$. The above
solutions should satisfy the asymptotics
$\varphi_{a,b}^{>}(x'\rightarrow\infty)=0$ and
$\varphi_{a,b}^{<}(x'\rightarrow-\infty)=0$. Matching the solutions
and the first derivatives at $x'=0$ gives a homogeneous set of
algebraic equations which in matrix form becomes
\begin{equation}\label{eq}
\left(
  \begin{array}{cccc}
    1 & 1 & -1 & -1 \\
    f_{+} & f_{-} & -g_{+} & -g_{-} \\
    \lambda_{+} & \lambda_{-} & \lambda_{+} & \lambda_{-} \\
    f_{+}\lambda_{+} & f_{-}\lambda_{-} & g_{+}\lambda_{+} & g_{-}\lambda_{-} \\
  \end{array}
\right)\left(
         \begin{array}{c}
           C_{1} \\
           C_{2} \\
           C_{3} \\
           C_{4} \\
         \end{array}
       \right)=0.
\end{equation}
Solutions are found when the determinant of the matrix is set to
zero from which we obtain the energy spectrum. Notice that Eq. (8)
leads to four solutions, of which two of them, i.e. $\pm u_{b}$, do
not satisfy Eqs. (\ref{eqSB0}) and are not acceptable. In the
limiting case $\epsilon\ll u_{b}$ we are able to obtain an
analytical expression for the energy,
\begin{equation}\label{eq}
\begin{split}
&\epsilon_{\pm}=\frac{u_{b}}{\alpha}\times\left\{4k'_{y}\sqrt{\epsilon_{0}}[u_{b}\sin(\theta/2) +k_{y}^{'2}\cos(\theta/2)]
\right.\\
&\left.\pm\big[56k_{y}^{'8}+14u_{b}^{4}+70u_{b}^{2}k_{y}^{'4}-k_{y}^{'2}
\epsilon_{0}(40k_{y}^{'4}+46{u_{b}}^{2})\big]^{1/2}~\right\}.
\end{split}
\end{equation}
where, $\epsilon_{0}=\sqrt{k_{y}^{'4}+{u_{b}}^{2}}$,
$\alpha=6k_{y}^{'4}+7{u}^{2}-6k_{y}^{'2}\epsilon_{0}$ and
$\theta=\tan^{-1}(u_{b}/k_{y}^{'2})$. Solving the above equation for
$\epsilon=0$ we find that $k'_{y}=\pm\sqrt{u_{b}/\sqrt{8}}$
($\approx 0.3$ for $u_{b}=0.25$).
\begin{figure}
\centering
\includegraphics[width=8cm]{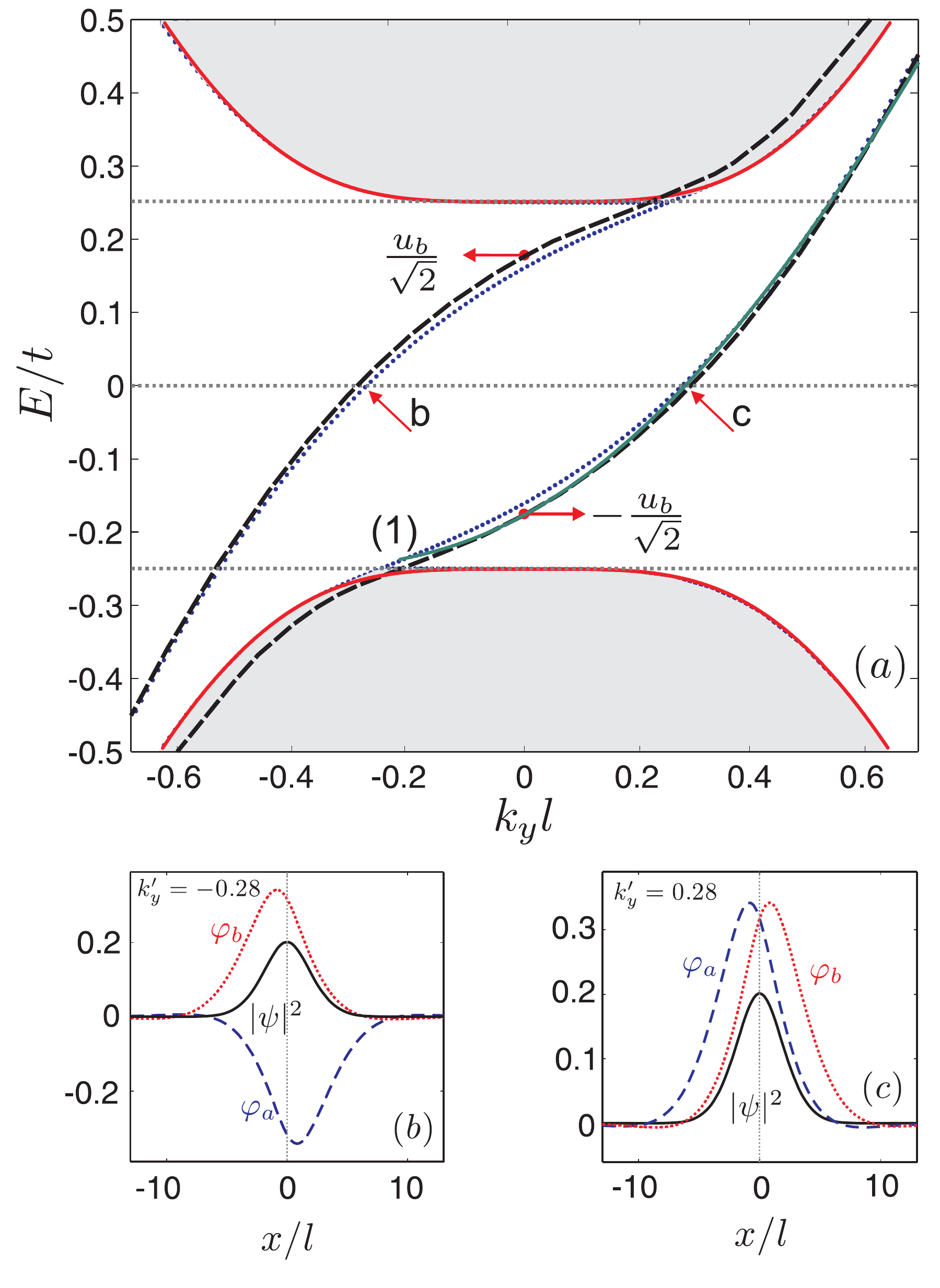}
\caption{(Color online) (a) Energy levels for a single kink profile
on bilayer graphene with $u_{b}=0.25$. Dotted curves are the
numerical results for $\delta=1$ and dashed curves are the
analytical results for $\delta=0$ using Eq. (9). The solid red
curves are the energy levels of a biased BLG. The solid green curve
(indicated by the symbol (1)) shows a fitted function to the
numerical results. The lower panels show the wave spinors and
probability density corresponding to the states that are indicated
by the arrows (b) and (c) in panel (a).} \label{fig1}
\end{figure}
\begin{figure}
\centering
\vspace*{1.0cm}
\hspace*{-1cm}
\includegraphics[width=8cm]{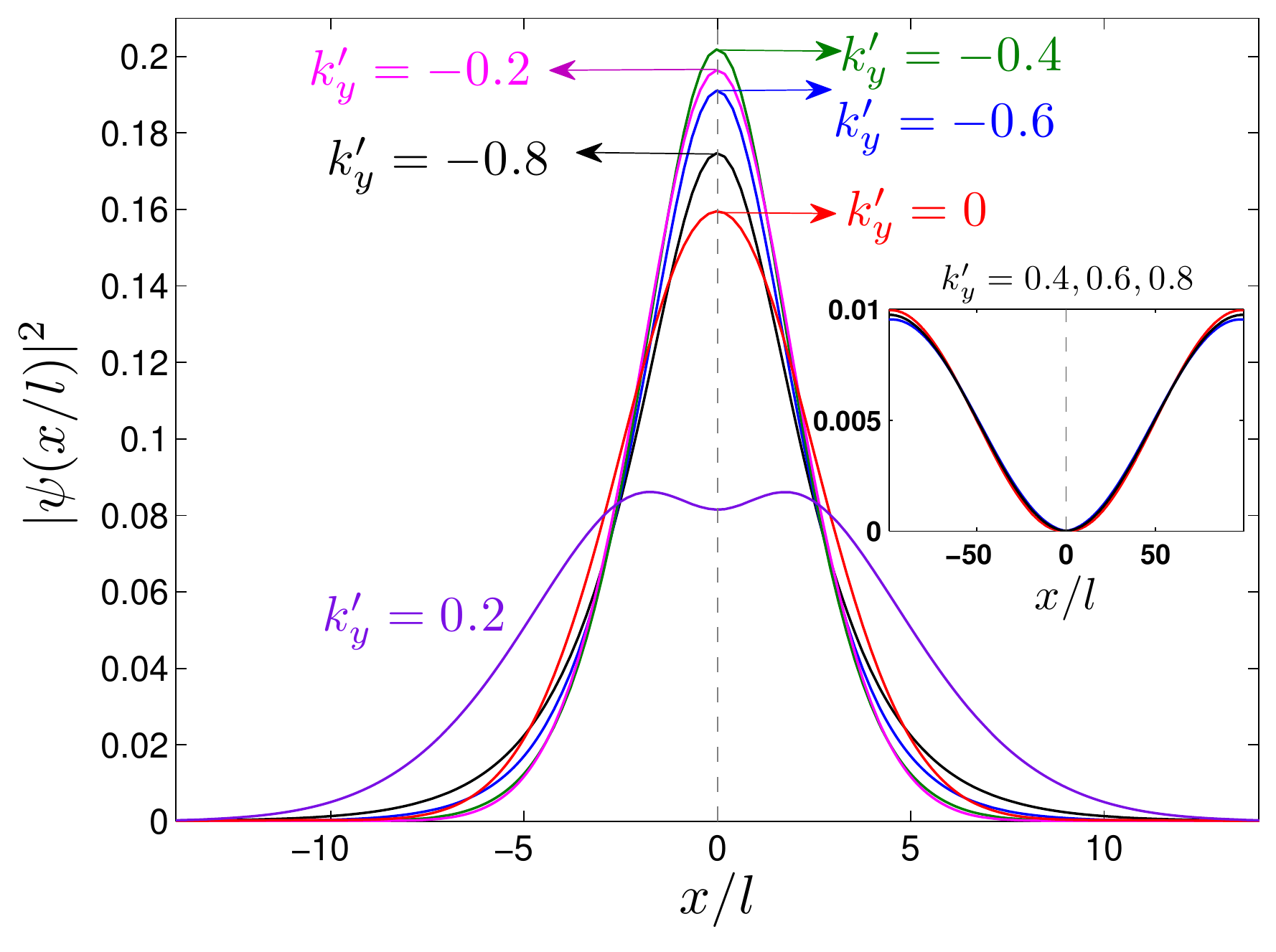}
\caption{(Color online) The probability densities of the topological
state in $k'_{y}=-0.8,-0.6,...$ which correspond to the state
indicated by $(1)$ in Fig. (\ref{fig1}). The inset shows the
probability density for $k'_{y}=0.4,0.6,0.8$} \label{Psi2S}
\end{figure}
\begin{figure}
\centering
\includegraphics[width=8cm]{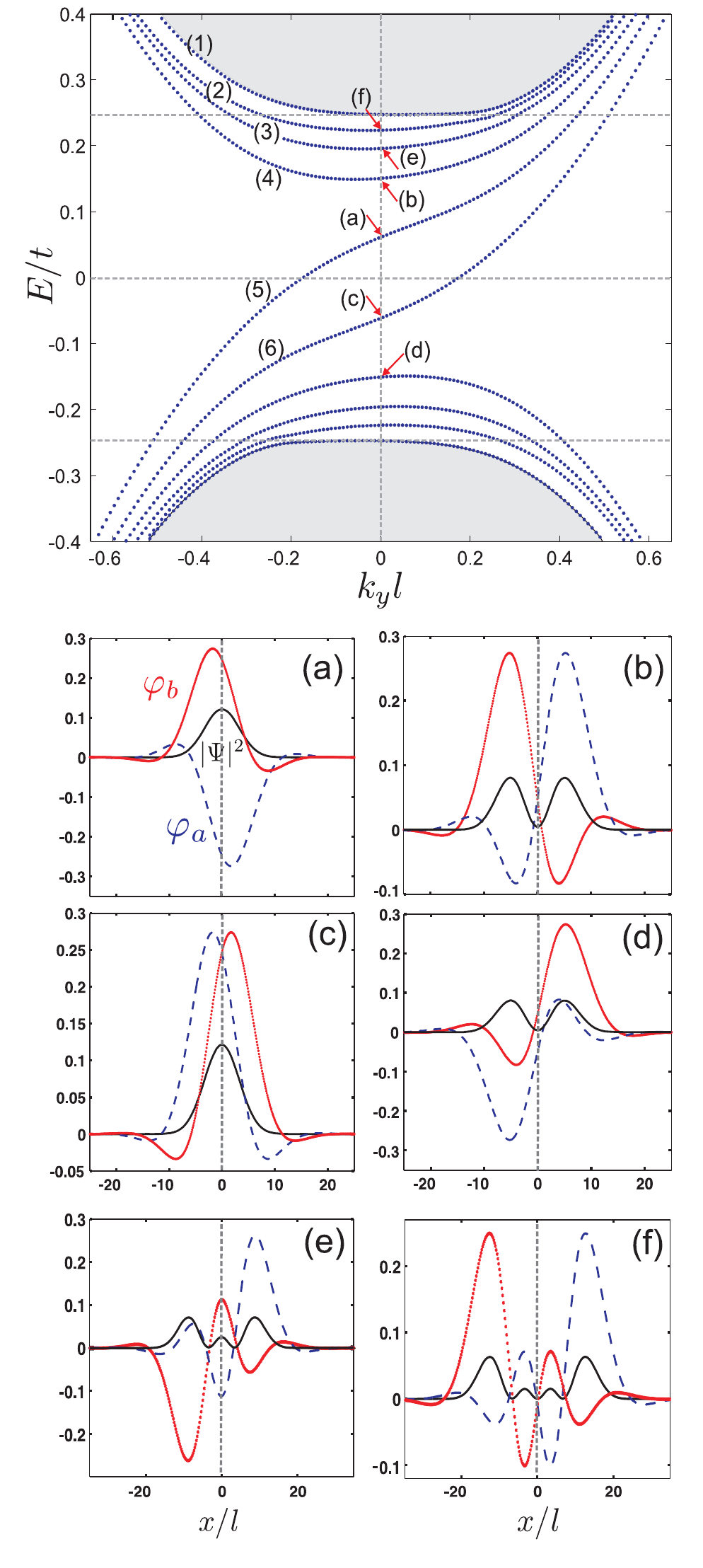}
\caption{(Color online) Upper panel: Energy levels for a single kink
profile in bilayer graphene with $u_{b}=0.25$ and $\delta=10$. The
energy states indicated by (a) and (c) are chiral states and those
indicated by (b),(d),(e) and (f) are the extra-bound states. Lower
panels: Real parts of the wave spinors and the corresponding
probability density for the two first electron and hole energy
levels at $k'_{y}=0$ as indicated in the upper panel.} \label{fig2}
\end{figure}
\begin{figure}
\centering
\includegraphics[width=8cm]{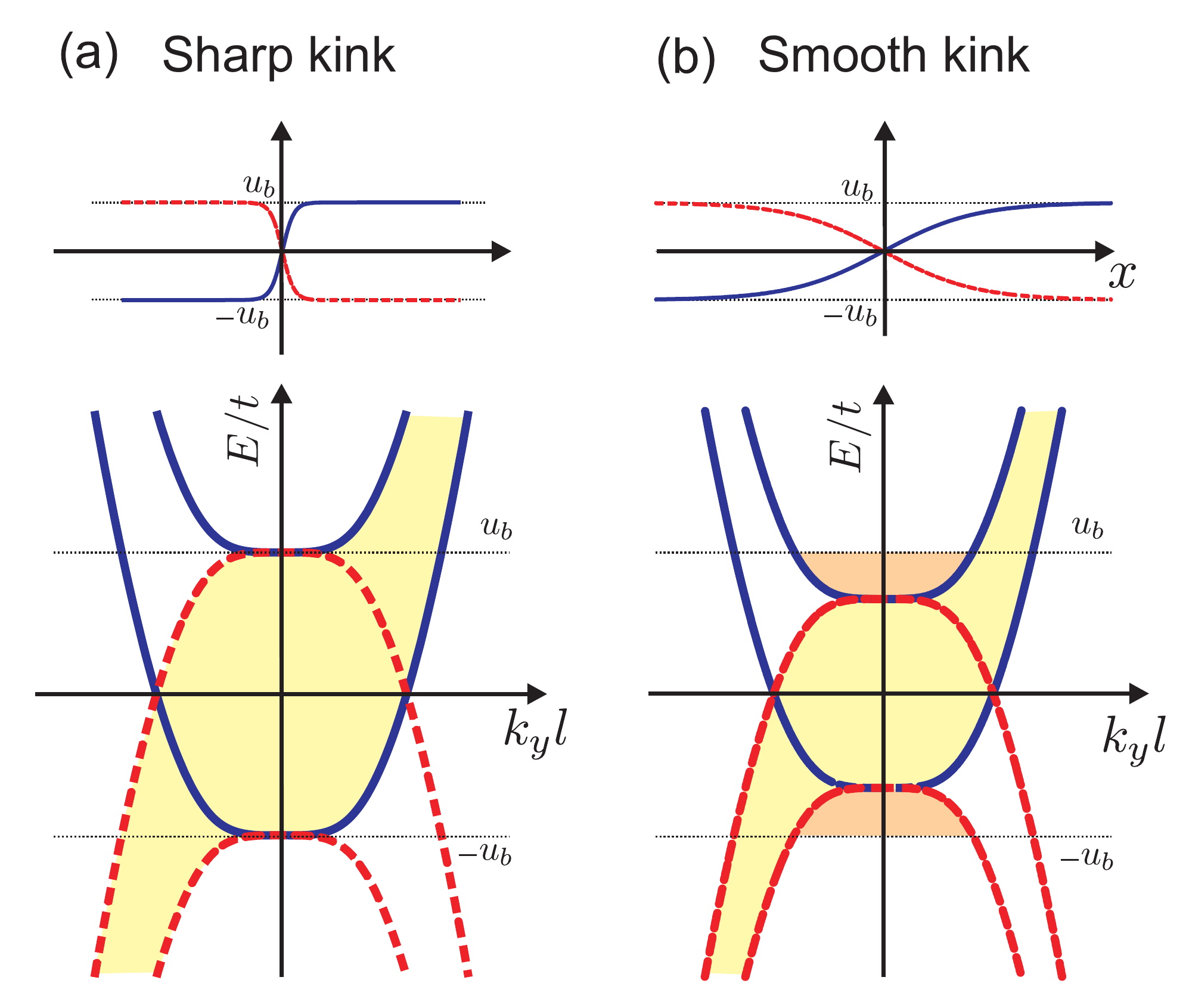}
\caption{(Color online) Upper panels: applied potential profile
$u(x)$ to the upper (solid blue curve) and the lower layer (red
dashed curve) for (a) sharp and (b) smooth kink profiles. Lower
panel: schematic representation of the energy spectrum in
($E,k_{y}$) space, corresponding to the blue and red dashed
potential profiles near the kink region. Topological states can be
found in the yellow region. For the smooth profile (b), extra bound
states can be found in the orange region.} \label{fig3}
\end{figure}

Next we consider a sharp kink potential in parallel with an antikink
potential which are located at $x'=-d$ and $x'=+d$. In this case we
have to consider three regions, i.e. $x'>d$ ($\psi_{I}(x')$),
$-d\leq x'\leq-d$ ($\psi_{II}(x')$) and $x'>d$ ($\psi_{III}(x')$)
and the solutions are given by
\begin{subequations}
\begin{eqnarray}
&&\psi_{I}(x')_{\pm}=\left(
               \begin{array}{c}
             e^{i\lambda_{\pm}x'} \\
             g_{\pm}e^{i\lambda_{\pm}x'}\\
               \end{array}
             \right), \\
&&\psi_{II}(x')_{\pm}=\left(
               \begin{array}{c}
             e^{\pm i\lambda_{\pm}x'}\\
             f_{\pm}e^{i\lambda_{\pm}x'}\\
               \end{array}
             \right),\\
&&\psi_{III}(x')_{\pm}=\left(
               \begin{array}{c}
             e^{-i\lambda_{\pm}x'} \\
             g_{\pm}e^{-i\lambda_{\pm}x'}\\
               \end{array}
             \right)
\end{eqnarray}
\end{subequations}
Matching the solutions and their first derivatives at $x'=\pm d$ leads to a set of eight algebraic equations which in matrix form becomes
\begin{widetext}
\begin{equation}\label{eq88}
\scriptstyle{
\left(
  \begin{array}{cccccccc}
    \kappa_{+}^{-} & \kappa_{-}^{-} & -\kappa_{+}^{-} & -\kappa_{-}^{-} & -\kappa_{+}^{+} & -\kappa_{-}^{-} & 0 & 0 \\
    g_{+}\kappa_{+}^{-} & g_{-}\kappa_{-}^{-} & -f_{+}\kappa_{+}^{-} & -f_{-}\kappa_{-}^{-} & -f_{+}\kappa_{+}^{+} & -f_{-}\kappa_{-}^{-} & 0 & 0 \\
     \lambda_{+}\kappa_{+}^{-} & \lambda_{-}\kappa_{-}^{-} & -\lambda_{+}\kappa_{+}^{-} & -\lambda_{+}\kappa_{-}^{-} & \lambda_{+}\kappa_{+}^{+} & \lambda_{-}\kappa_{-}^{-} & 0 & 0 \\
    g_{+}\lambda_{+}\kappa_{+}^{-} & g_{-}\lambda_{-}\kappa_{-}^{-} & -f_{+}\lambda_{+}\kappa_{+}^{-} & -f_{-}\lambda_{-}\kappa_{-}^{-} & f_{+}\lambda_{+}\kappa_{+}^{+} & f_{-}\lambda_{-}\kappa_{-}^{-} & 0 & 0 \\
    0 & 0 &  \kappa_{+}^{+} & \kappa_{-}^{+} & \kappa_{+}^{-} & \kappa_{-}^{-} & -\kappa_{+}^{-} & -\kappa_{-}^{-} \\
    0 & 0 &  f_{+}\kappa_{+}^{+} & f_{-}\kappa_{-}^{+} & f_{+}\kappa_{+}^{-} & f_{-}\kappa_{-}^{-} & -g_{+}\kappa_{+}^{-} & -g_{-}\kappa_{-}^{-} \\
    0 & 0 & \lambda_{+}\kappa_{+}^{-} & \lambda_{-}\kappa_{-}^{+} & -\lambda_{+}\kappa_{+}^{-} & -\lambda_{-}\kappa_{-}^{-} & \lambda_{+}\kappa_{+}^{-} & \lambda_{-}\kappa_{-}^{-} \\
    0 & 0 & f_{+}\lambda_{+}\kappa_{+}^{+} & f_{-}\lambda_{-}\kappa_{-}^{+} & -f_{+}\lambda_{+}\kappa_{+}^{-} & -f_{-}\lambda_{-}\kappa_{-}^{-} & g_{+}\lambda_{+}\kappa_{+}^{-} & g_{-}\lambda_{-}\kappa_{-}^{-} \\
  \end{array}
\right)\left(
         \begin{array}{c}
           C_{1} \\
           C_{2} \\
           C_{3} \\
           C_{4} \\
           C_{5} \\
           C_{6} \\
           C_{7} \\
           C_{8} \\
         \end{array}
       \right)=\boldsymbol{0}.}
\end{equation}
\end{widetext}
where, $\kappa_{\pm}^{+}=\exp(i\lambda_{\pm}d)$ and
$\kappa_{\pm}^{-}=\exp(-i\lambda_{\pm}d)$. Setting the determinant
to zero gives the energy spectrum.
\begin{figure}
\centering
\includegraphics[width=8cm]{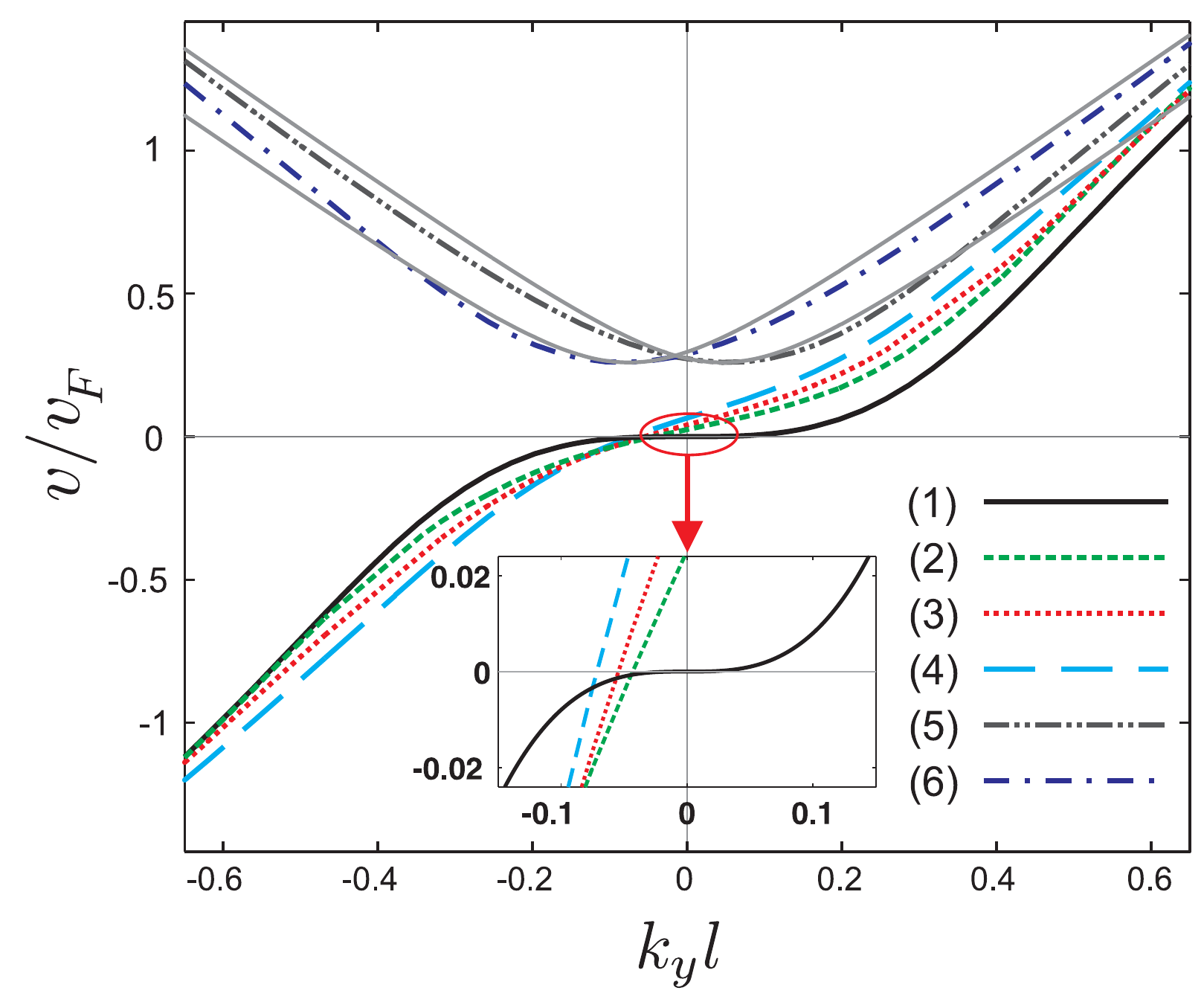}
\caption{(Color online) The carrier velocity in the single kink
profile for the energy levels which are indicated by $(1), (2),...$
in Fig. \ref{fig2}. The gray solid curves are the fitted functions
to curve (5) and (6).} \label{fig4}
\end{figure}

\section{Single kink}
\begin{figure}
\centering
\includegraphics[width=8cm]{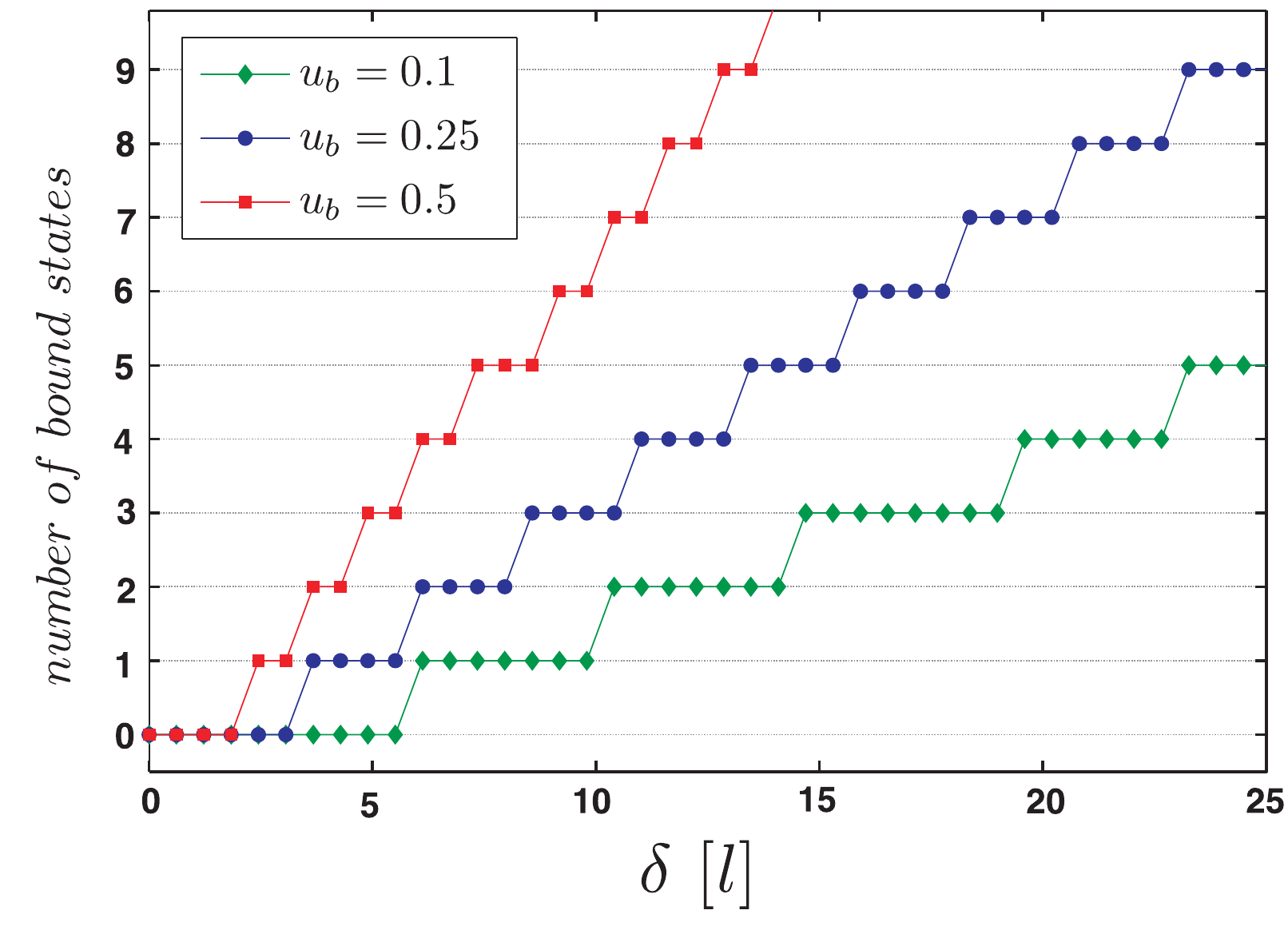}
\caption{(Color online) Number of additional bound states as
function of the width of the interface $\delta$ for $u_{b}=0.1$,
$0.25$ and $0.5$.} \label{fig5}
\end{figure}
\begin{figure}
\centering
\includegraphics[width=8.5cm]{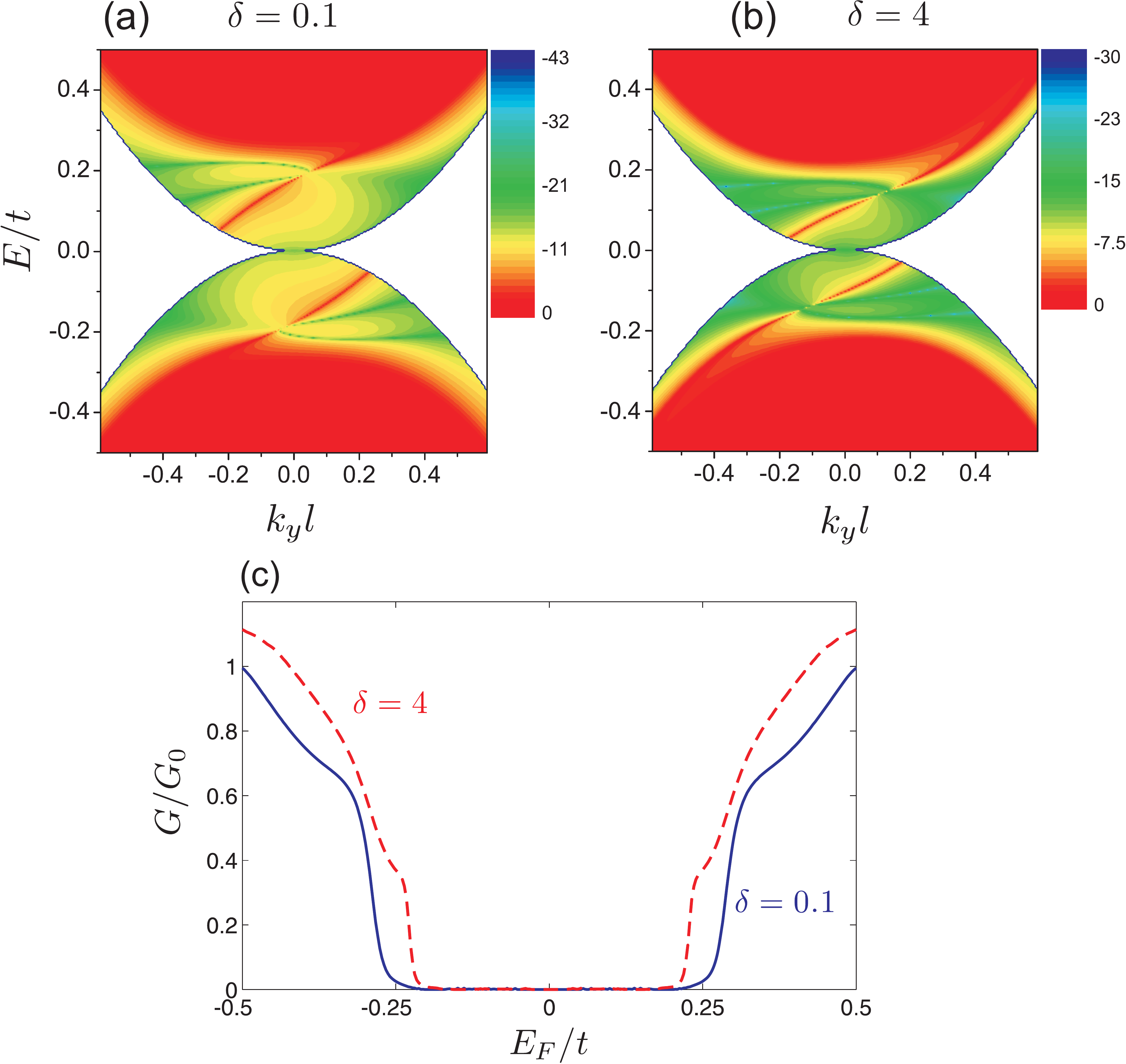}
\caption{(Color online) (a,b) Contour plot of the transmission $T$
(in logarithmic scale) through a single kink structure with
$L_{x}=20$ ($\approx 32~nm$), $u_{b}=0.25$ for (a) $\delta=0.1$ and
(b) $\delta=4$. (c) Conductance $G/G_{0}$ vs Fermi energy of a
single kink profile for the same parameters as (a,b).}\label{CS}
\end{figure}
\subsection{Influence of the smoothness of the kink profile}
In the general case of $\delta\neq0$ we solve the set of second
order differential Eqs. \ref{eqS}(a,b) numerically, using the finite
difference technique. Figure \ref{fig1}(a) shows the spectrum for a
single potential kink as function of the wavevector along the kink
for zero magnetic field. We consider a relatively sharp kink, i.e.
$\delta=1$, and compare the numerical results with the analytical
solution (dashed black curves) form Eq. (9) for the case of a sharp
profile ($\delta=0$). The shaded region corresponds to the continuum
of free states. The solid red curves correspond to the energy levels
of a biased BLG which can be obtained using Eq. (1) as,
\begin{equation}\label{eqbBLG}
\displaystyle{\epsilon=\pm\sqrt{k_{y}^{'4}+u_{b}^{2}}}
\end{equation}
The dotted horizontal lines correspond to $\epsilon = \pm u_b=\pm
0.25$ and $\epsilon = 0$. These results are valid in the vicinity of
a single valley ($K$) and show that the topological states have a
unidirectional character of propagation, i.e. they are chiral
states\cite{Morpurgo}, with positive group velocity. The topological
levels can be fitted to
$\displaystyle{\epsilon=a\sqrt{(k'_{y}-k_{0})^{4}+(u_{b}/a)^{2}}}-1/2$
with $a=0.5$ and $k_{0}=-0.65$ being the fitting parameters (see
green solid curve). For localized states around the $K'$ valley, we
have $E_{K'}(k_{y}) = -E_K(k_{y})$ and the charge carriers move in
the opposite direction. In order to consider the energy levels for
the $K'$ valley, $u(x)$ in Eqs. (3) should be replaced with $-u(x)$.
Then using the transformations $\epsilon\rightarrow-\epsilon$ and
$\varphi_{a}\rightarrow-\varphi_{a}$ (or
$\varphi_{b}\rightarrow-\varphi_{b}$) leads to the same equation as
for the $K$ valley. Thus the $E_K(k)=-E_K'(k)$ symmetry remains even
in the presence of an uniform perpendicular magnetic field (i.e.
$\beta\neq 0$). Notice that the wavespinors corresponding to the $K$
and the $K'$ valleys are related to each other by
$\varphi_{a}^{K}\rightarrow-\varphi_{a}^{K'}$ (or
$\varphi_{b}^{K}\rightarrow-\varphi_{b}^{K'}$ ) while the sign of
the other component does not change.

Panels (b) and (c) of Fig. \ref{fig1} present the real parts of the
spinor components and the probability density for the states
indicated by the arrows (b) and (a) in panel (a), corresponding to
$k'_y = -0.28$ (b) and $k'_y = 0.28$ (c). These electron states are
localized at the position of the potential kink. Notice that the
solutions of Eqs. (3) are related by the transformations $\varphi_a
\rightarrow -\varphi_b$, $\varphi_b \rightarrow \varphi_a$, $k'_y
\rightarrow -k'_y$ and $\epsilon \rightarrow -\epsilon$ and
consequently for $\epsilon=0$ the solutions in Figs. \ref{fig1}(b)
and \ref{fig1}(c) have the same probability distribution. For the
case of $k'_{y}=0$ the solutions of Eq. (9) are $\epsilon_{\pm}=\pm
u_{b}/\sqrt{2}$ which result in the following wavespinors,
\begin{subequations}\label{eqk0}
\begin{eqnarray}
&&\varphi_{a}^{<}=(1\mp\sqrt{2})[e^{-i(\lambda_{+}x'+\frac{\pi}{4})}\mp e^{i(\lambda_{-}x'+\frac{\pi}{2})}], \\
&&\varphi_{a}^{>}=e^{i(\lambda_{+}x'+\frac{\pi}{4})}\mp  e^{-i\lambda_{-}x'}, \\
&&\varphi_{b}^{<}=-\frac{(1\mp\sqrt{2})}{(\epsilon_{\pm}-u_{b})}\big[\lambda_{+}^2 e^{-i(\lambda_{+}x'+\frac{\pi}{4})}\mp\lambda_{-}^2e^{i(\lambda_{-}x'
              +\frac{\pi}{2})}\big],~~~~~~\\
&&\varphi_{b}^{>}=\frac{-1}{\epsilon_{\pm}+u_{b}}\big[\lambda_{+}^2 e^{i(\lambda_{+}x'+\frac{\pi}{4})}\mp \lambda_{-}^2e^{-i\lambda_{-}x'}\big].
\end{eqnarray}
\end{subequations}
where, $\displaystyle{\lambda_{\pm}=\gamma(1\pm i)}$ with $\gamma$
being $\sqrt{u_{b}/8}$. Notice that in the above equations $e^{\pm
i\lambda_{\pm}x'}$ leads to an oscillating contribution $e^{\pm
i\Re(\lambda_{\pm})x'}$ with an evanescent $e^{\mp
\Im(\lambda_{\pm})x'}$ part. The oscillating part is strongly damped
and therefore Eqs. (\ref{eqk0}) corresponds to localized
wavespinors. Expanding Eqs. (\ref{eqk0}) around $x'=0$ we obtain for
the second derivative of the wavespinors, 
\begin{subequations}\label{eq}
\begin{eqnarray}
&&\frac{\partial^{2}}{\partial x'^2}\Re[\varphi_{a}^{<}(x'\rightarrow 0)]=(1\pm\sqrt{2})(2+\sqrt{2})\gamma,~~~~\\
&&\frac{\partial^{2}}{\partial x'^2}\Re[\varphi_{b}^{<}(x'\rightarrow 0)]= \frac{(4-2\sqrt{2})\gamma^3}{\epsilon_{\pm}-u_{b}}<0.~~~~~
\end{eqnarray}
\end{subequations}
This indicates that $\Re[\varphi_{a}](\Re[\varphi_{b}])$ has its
maximum value located at $x'<0$($x'>0$) for
$\epsilon_{+}=u_{b}/\sqrt{2}$ while the opposite is found for
$\epsilon_{-}=-u_{b}/\sqrt{2}$ which is also evident from Figs.
\ref{fig1}(b,c).

In Fig. \ref{Psi2S} we show the probability densities corresponding
to one of the topological branches (for the state which is labeled
by $(1)$ in Fig. \ref{fig1}(a)) at several $k'_{y}$ values. As shown
in the inset of Fig. \ref{Psi2S} for those $k'_{y}$ values where the
topological state merged with the continuum spectrum the carries are
no longer  confined by the kink potential.

Next we increase the smoothness of the kink potential and
investigate how the energy spectrum changes. In Fig. \ref{fig2} the
energy levels as function of $k_{y}$ are shown for the smooth kink
profile $\delta=10$, where in addition to the chiral states several
branches are seen which are split off from the continuum. In order
to understand the physical origin of those new states we show in the
lower panels of Fig. \ref{fig3} a cartoon of the low energy spectra
for the (a) sharp and (b) smooth profiles where, the chiral states
appear in the yellow regions and those additional states are found
in the orange region. Increasing the smoothness of the kink
potential leads to the creation of a region below the energy gap
which allows for carriers to be confined near the kink. Therefore,
extra bound states can be created in the orange region (lower panel
in Fig. \ref{fig3}(b)). The wavefunctions for $k_{y}=0$ of the two
chiral states and the new bound states are shown in the lower panels
of Fig. \ref{fig2}. The new bound states are also bound in the
$x$-direction near $x=0$ but the electron states are more extended
and have a clear nodal character near $x=0$.

Figure \ref{fig4} shows the velocity of the carriers for the states
which are indicated by $(1),(2),...$ in Fig. \ref{fig2}. The chiral
states ((5),(6)) are only shown for the $K$ valley and they have
positive velocity. The curves (5),(6) can be fitted to
$v/v_{F}\approx a\sqrt{(k'_{y}-k'_{0})^2+b^2}$ (see the solid gray
curves) with $a=1.8, b=0.15$ being the fitting parameters and
$k'_{0}=\pm 0.08$ corresponds to the minimum point in the curves (5)
and (6). Notice that the extra bound states $((2),(3),(4))$ have a
slightly nonzero velocity at $k'_{y}=0$ which is a consequence of
the asymmetric energy dispersion as seen in Fig. \ref{fig2}. Curve
$(1)$ corresponds to the energy spectrum of a biased BLG which is
given by Eq. (\ref{eqbBLG}) and results in the velocity
$\displaystyle{v/v_{F}=[\partial\epsilon/\partial
k'_{y}]=2k_{y}^{'3}/\sqrt{k_{y}^{'4}+u_{b}^{2}}}$ which is zero for
$k'_{y}=0$ in a biased BLG (black solid curve in Fig. \ref{fig4}).

As mentioned before for smooth kink potentials additional 1D bound
states appear and the number of these bound states can be related to
the height of the gate voltage $u_{b}$ and the smoothness($\delta$)
at the interface. Figure \ref{fig5} shows the number of these extra
bound states for three different $u_{b}$ values as function of the
width $\delta$. The first bound state for $u_{b}=0.1, 0.25, 0.5$
appears respectively at $\delta\approx 6, 4, 2$ in the absence of
magnetic field. Notice also that for fixed $\delta$ the number of
extra bound states increases with $u_{b}$ in agreement with the
qualitative picture shown in Fig. \ref{fig3}(b).

We also calculate the transmission of an electron through the kink
structure in a system of size $-L_{x}/2<x<L_{x}/2$ and
$-L_{y}/2<y<L_{y}/2$. No bias nor magnetic field is assumed in the
$x<-L_{x}/2$ and $x>L_{x}/2$ regions. We assume that $L_{y}>>L_{x}$
and the electrons are free to move in the $y$-direction whereas,
they are confined in the $x$-direction. Associated with each real
$\lambda_{\pm}$ there are two right(left) propagating modes,
$\psi^{>}_{\pm}(\psi^{<}_{\pm})$ which are given by Eqs.
(\ref{eqi}). In the region $I$ ($x<-L_{x}/2$) two incident
right-traveling modes $\psi^{>}_{\pm}$ can be reflected into two
left-traveling modes $\psi^{<}_{\pm}$,
\begin{equation}\label{eqR}
\Psi^{I}_{\pm}=\psi^{>}_{\pm}+r_{\pm}^{+}\psi^{<}_{+}+r_{\pm}^{-}\psi^{<}_{-}.
\end{equation}
where, $t_{\pm}^{\pm}$ ($r_{\pm}^{\pm}$) are the transmission
(reflection) amplitudes. The propagating modes in region $I$ can
also be transmitted to region $III$ ($x>L_{x}/2$) in the
right-traveling modes,
\begin{equation}\label{eqT}
\Psi^{III}_{\pm}=t_{\pm}^{+}\psi^{>}_{+}+t_{\pm}^{-}\psi^{>}_{-}.
\end{equation}
The wevefunctions in regions $I$ and $III$ can be connected by the
transfer matrix $M$ where at the kink-potential boundaries we have
\begin{figure}
\centering
\includegraphics[width=8cm]{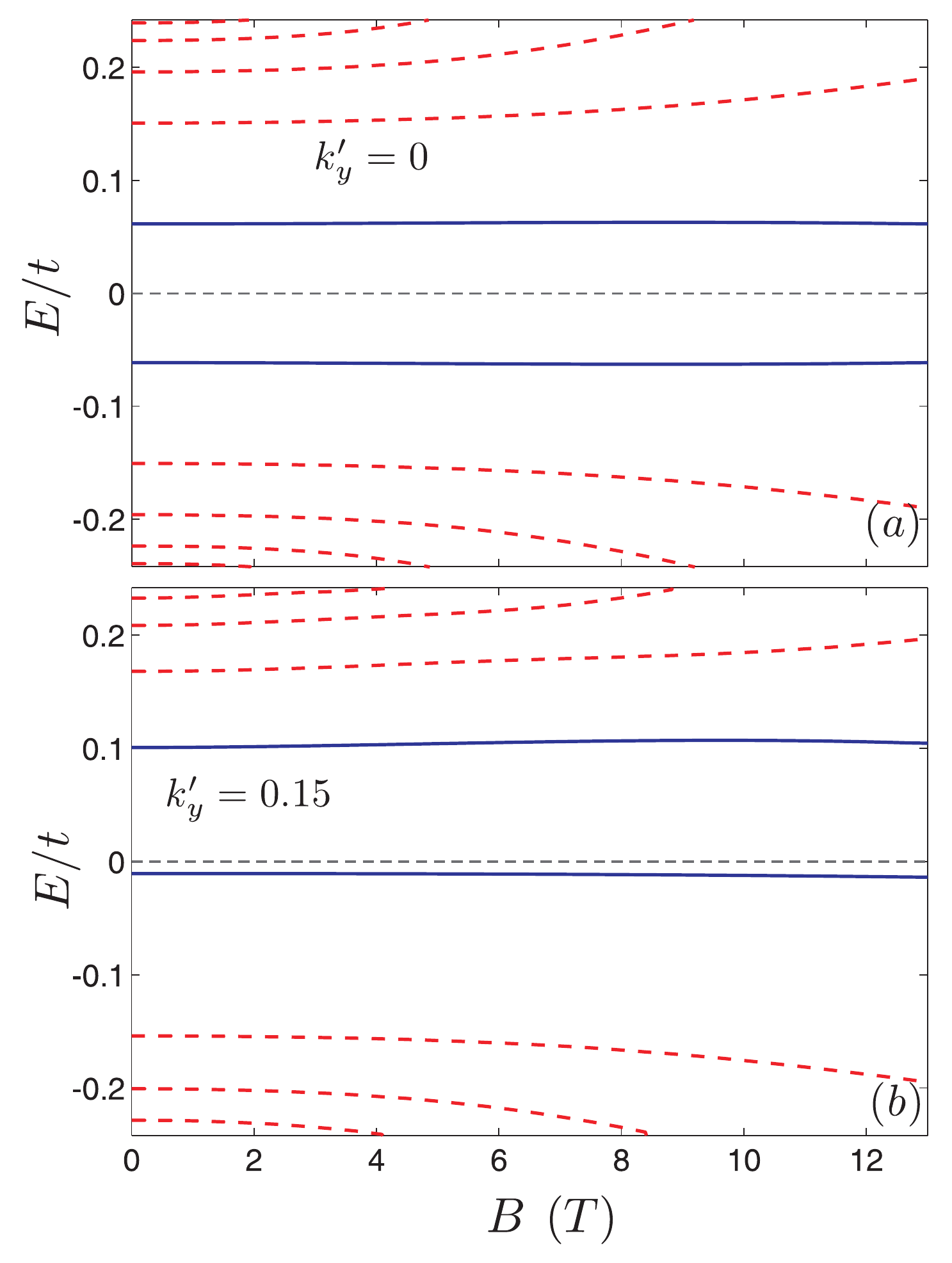}
\caption{(Color online) Energy levels of a single potential kink
profile as function of the external magnetic field with $u_{b}=0.25$
and $\delta=10$ for (a) $k_{y}l=0$ and (b) $k_{y}l=0.15$. The full
blue curves are the topological states and the dashed red curves are
the extra bound states.} \label{fig6}
\end{figure}
\begin{figure}
\centering
\includegraphics[width=8cm]{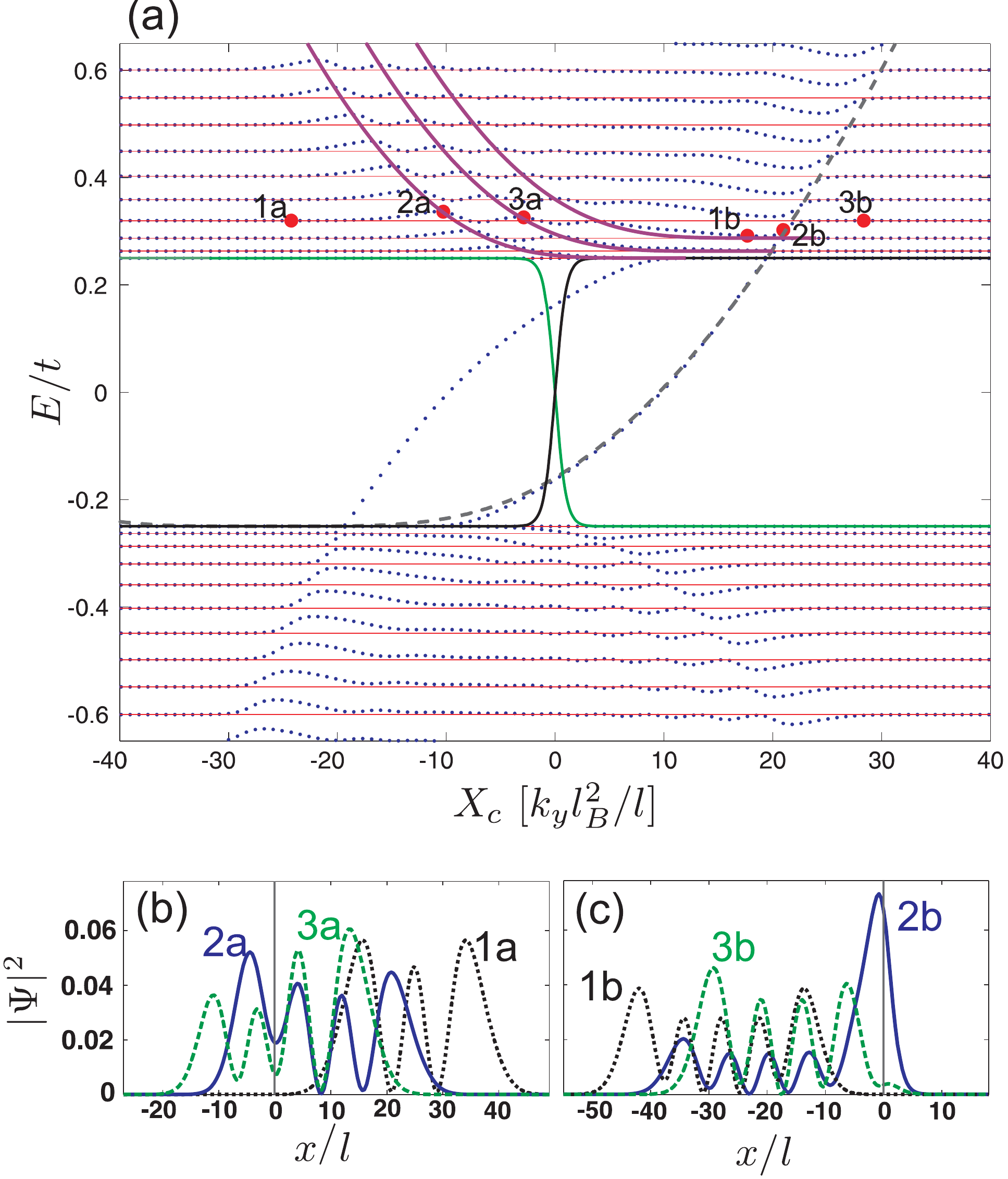}
\caption{(Color online) (a) Energy spectrum of a single kink profile
in bilayer graphene as function of the cyclotron orbit coordinate
$X_{c}$ for $B=7~T$, $u_{b}=0.25$ and $\delta=1$. The dashed gray
curve shows a fitted function to the numerical results. The solid
black and green curves are respectively the potential in the upper
and lower layer. Solid red lines corresponds to the Landau levels of
a biased BLG. The solid purple curves show fitted functions (given
in the text) to the position of the resonances. (b,c) The
probability densities for the indicated points by red full circles
in the spectrum.}\label{Xcs1}
\end{figure}
\begin{figure}
\centering
\includegraphics[width=8cm]{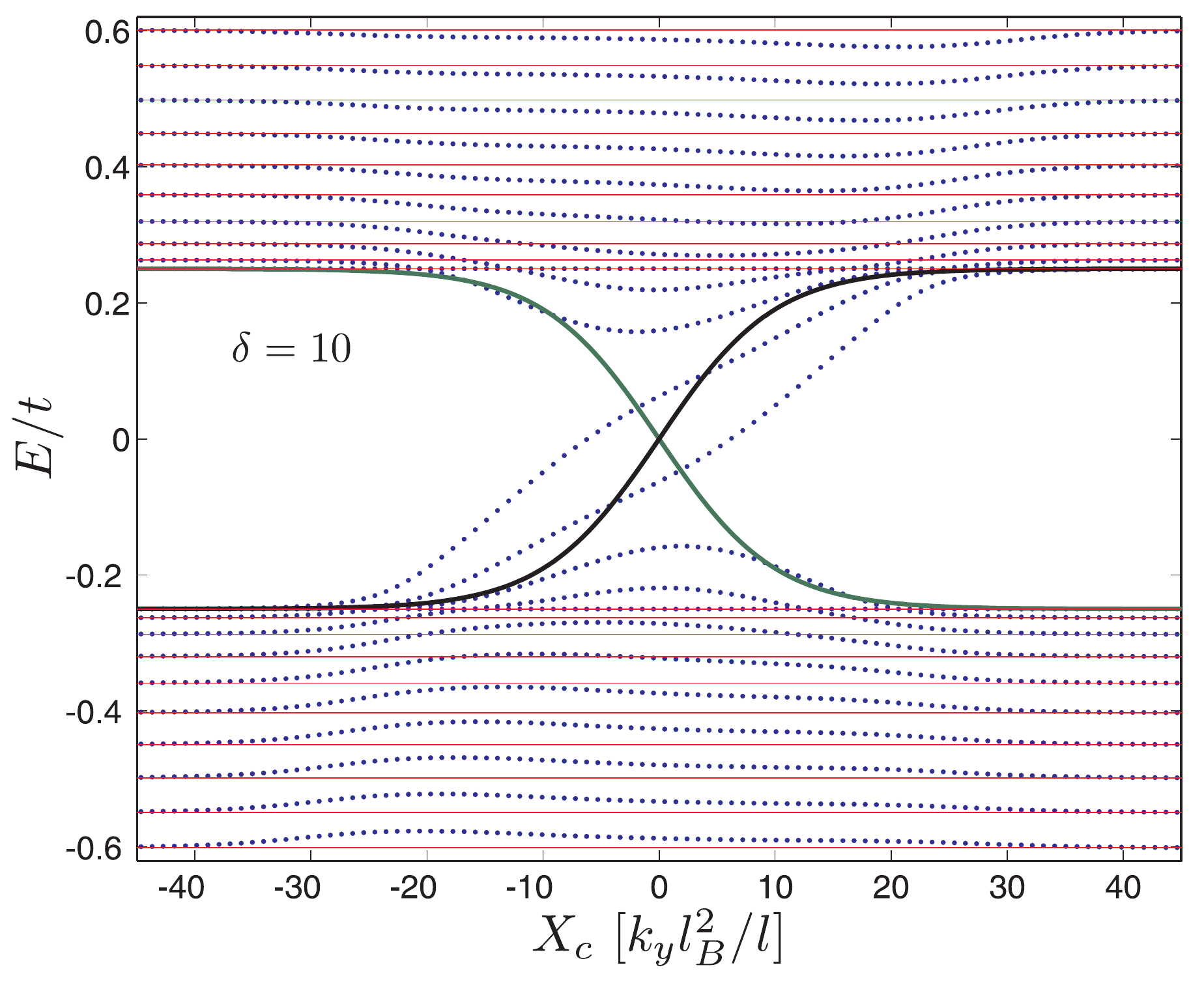}
\caption{(Color online) The same as Fig. \ref{Xcs1} but for
$\delta=10$. The solid black and green curves are the potential,
respectively, in the upper and lower layer.}\label{Xcs10}
\end{figure}
\begin{figure}
\centering
\includegraphics[width=8cm]{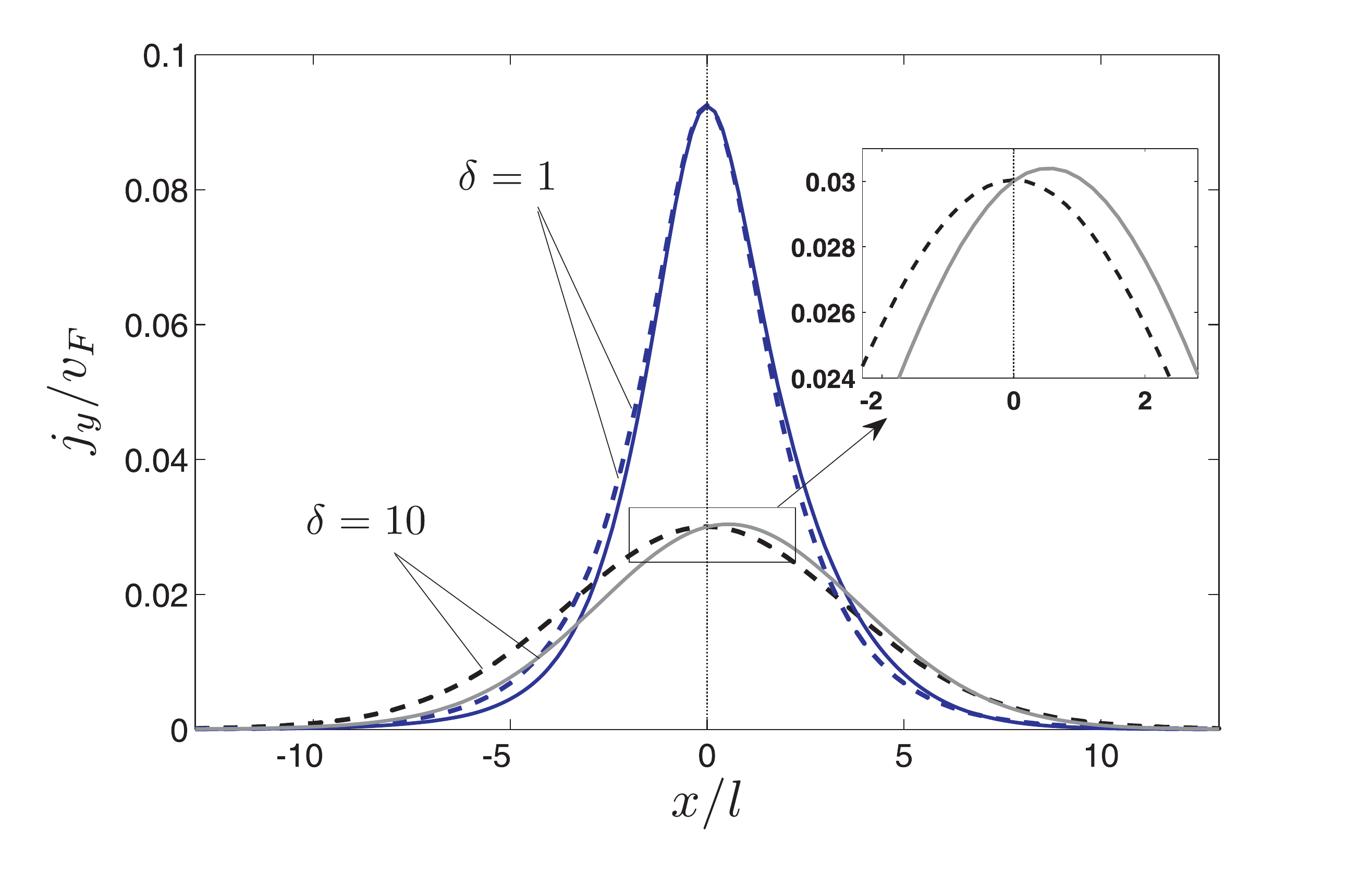}
\caption{(Color online) $y$-component of the persistent current
corresponding to the topological state of a single kink potential as
function of the $x$ direction for zero magnetic field (dashed
curves) and $B=5~T$ (solid curves) with $k'_{y}=0$ and $u_{b}=0.25$.
Blue curves display the current density for $\delta=1$ and black
curves are the corresponding results for $\delta=10$.} \label{jyS}
\end{figure}
\begin{figure}
\centering
\includegraphics[width=8cm]{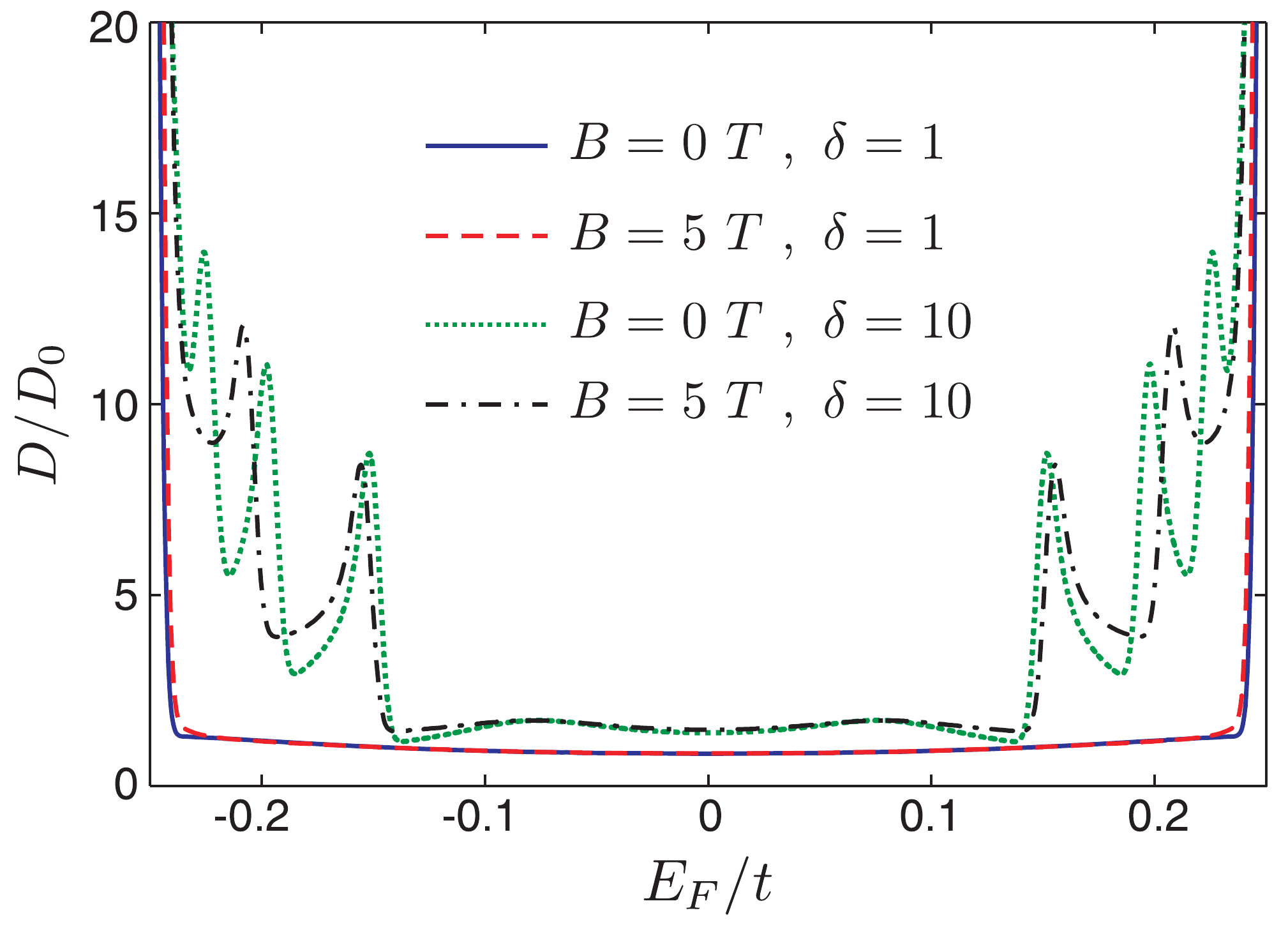}
\caption{(Color online) Density of states (DOS) of a single kink
potential for $\delta=1$ (solid and dashed curves) and $\delta=10$
(dotted and dashed-dotted curves) with $u_{b}=0.25$ for two
different magnetic field values.}\label{DOSs}
\end{figure}
\begin{figure}
\centering
\includegraphics[width=8cm]{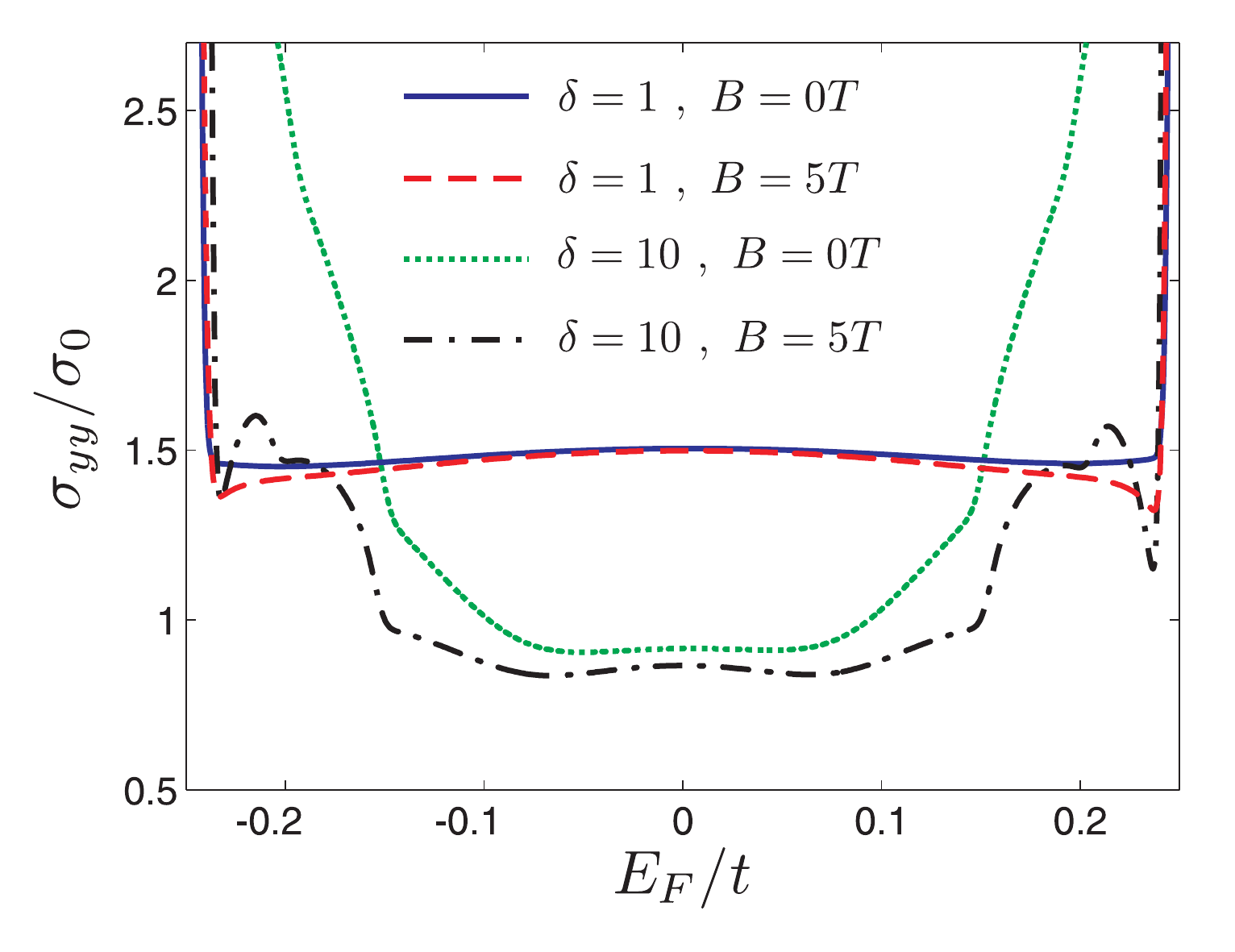}
\caption{(Color online) Conductivity $\sigma_{yy}$ as function of
the Fermi energy $E_{F}$ for a single kink potential with
$u_{b}=0.25$. Other parameters are indicated in the
figure.}\label{sigmaS}
\end{figure}
\begin{equation}\label{eqtransfer}
\Psi^{I}_{\pm}(-L_{x}/2)=M\Psi^{III}_{\pm}(L_{x}/2).
\end{equation}
The transmission (or reflection) amplitude can be found by
substituting Eqs. (\ref{eqR}) and (\ref{eqT}) in the above equation.
The four transmission amplitudes $t_{\pm}$ for given $\epsilon$ and
$k_y$ can be combined in the transmission matrix
\begin{equation}\label{eqTM}
t(\epsilon,k_{y})=\left(
                    \begin{array}{cc}
                      t_{+}^{+} & t_{-}^{+} \\
                      t_{-}^{+} & t_{-}^{-} \\
                    \end{array}
                  \right).
\end{equation}
The total transmission amplitude $T$ is given by\cite{snyman}
$T=Tr(tt^{\dag})$. The two-terminal conductance of such an
asymmetric potential profile in bilayer graphene can be calculated
using the Landauer formula which is given by\cite{ramezani,michael}
\begin{equation}
G=G_{0}\int T(E_{F},k'_{y})dk'_{y}
\end{equation}

\begin{figure*}
\centering
\includegraphics[width=12cm]{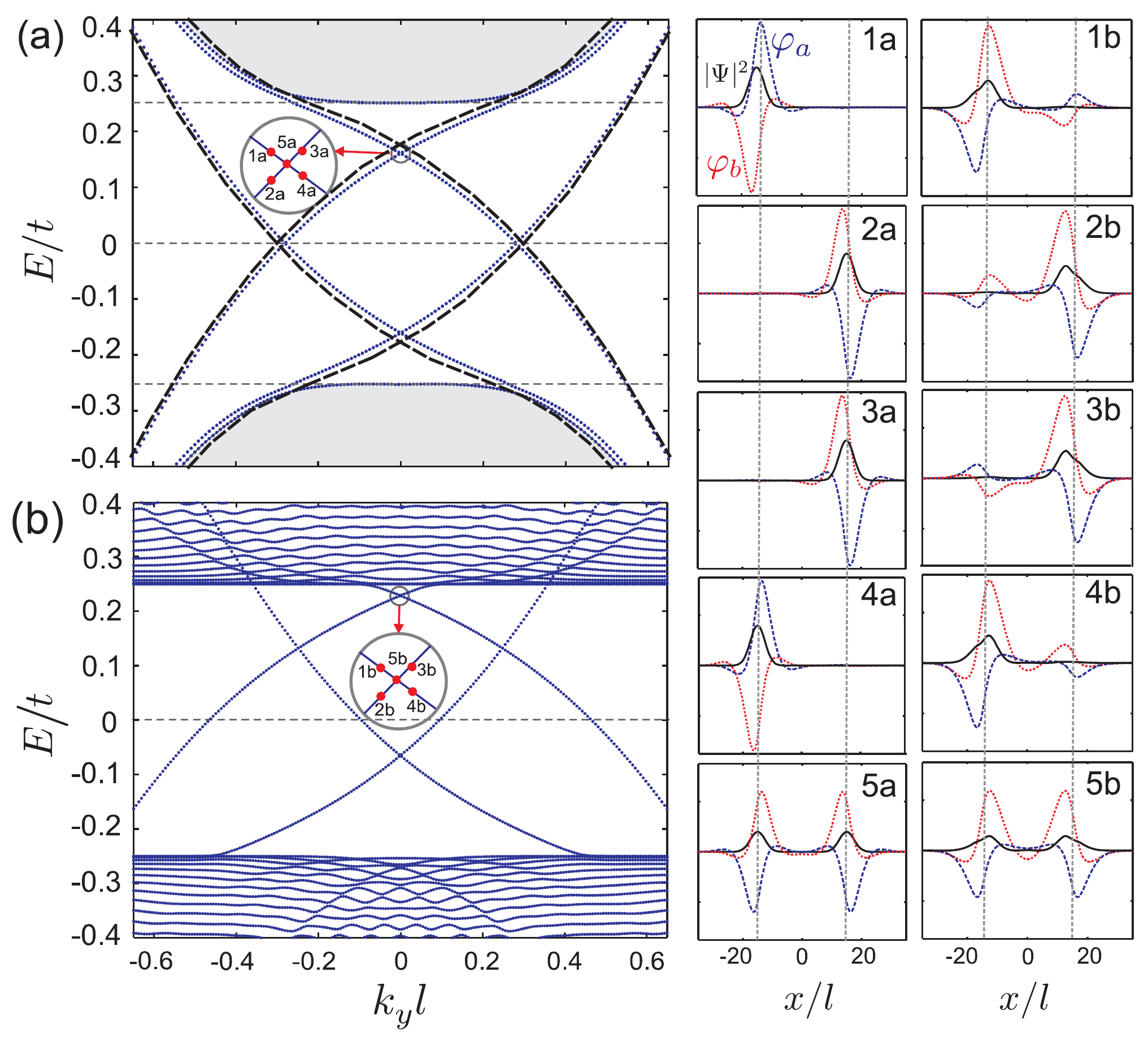}
\caption{(Color online) Left panels: Energy levels of a
kink-antikink profile in bilayer graphene with $u_{b}=0.25$,
$\delta=1$, $d=15$ for (a) $B=0~T$ and (b) $B=3~T$. The black dashed
curves display analytical results as obtained from Eq. (11). Right
panels: Real part of the wavespinors and the corresponding
probability density for the points that are indicated in the panels
(a,b).} \label{fig8}
\end{figure*}
\begin{figure*}
\centering
\includegraphics[width=12 cm]{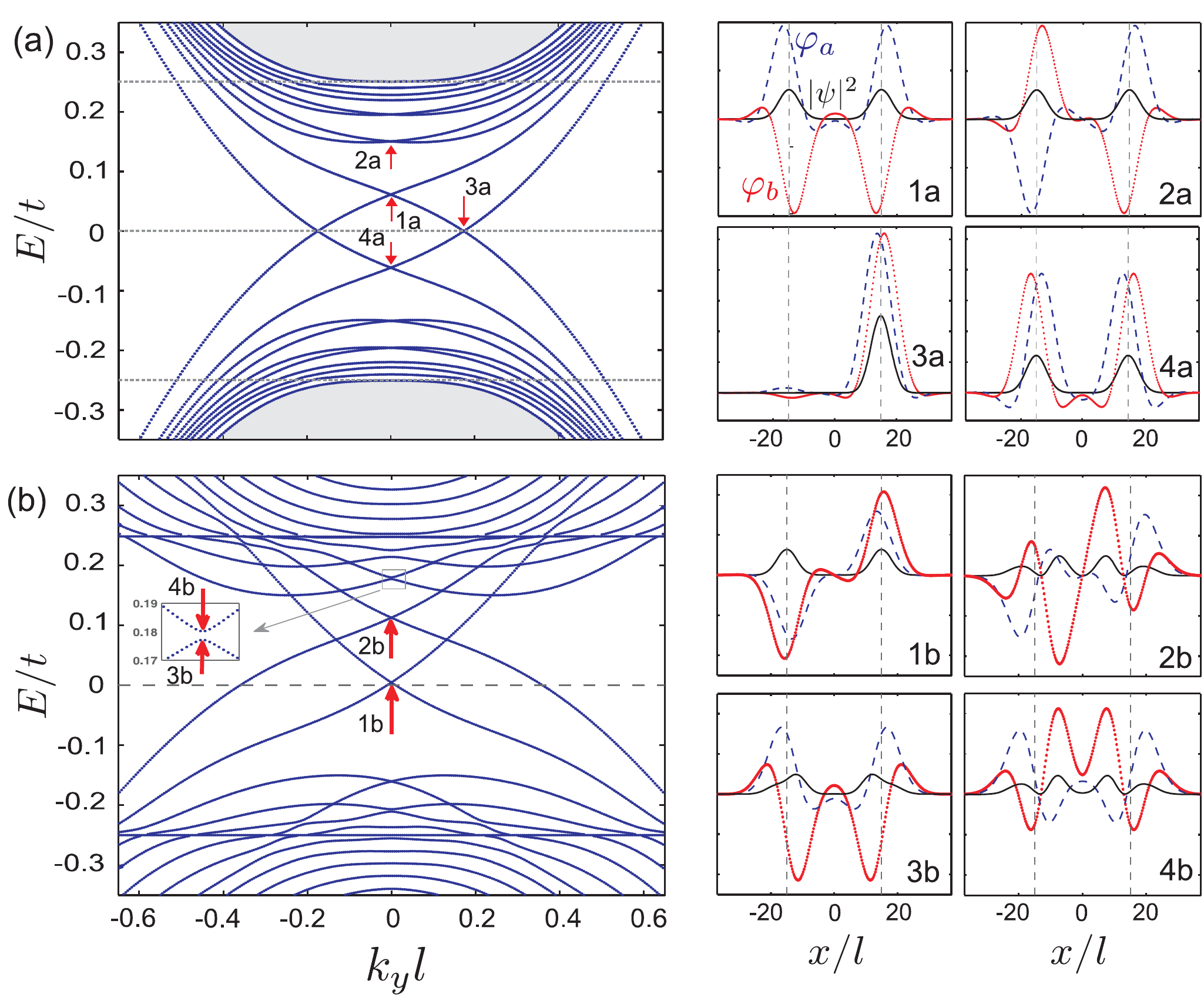}
\caption{(Color online) The same as Fig. \ref{fig8} but for a smooth
kink-antikink profile with $\delta=10$.} \label{fig11}
\end{figure*}

Here, $\displaystyle{G_{0}=(2e^{2}/h)(L_{y}|E_{F}|/\pi \hbar
v_{F})}$ is the conductance unit per valley and per spin. In Figs.
\ref{CS}(a,b) we show a contour plot of the transmission probability
in logarithmic scale for the kink structure with $L_{x}=20$ (in
dimensionless unit). The transmission probability has the symmetry
$T(k_{y},E)=T(-k_{y},-E)$. The conductance as function of the Fermi
energy for the single kink profile is shown in panel (c) for
$\delta=0.1$ (blue solid curve) and $\delta=4$ (red dashed curve).
For the case $\delta=4$ the smoothness of the potential at $x=0$
leads to a higher transmittance and consequently a higher
conductance around $\epsilon\approx u_{b}$ (see panel (b) and the
dashed curve in panel (c)).
\begin{figure*}
\centering
\includegraphics[width=12cm]{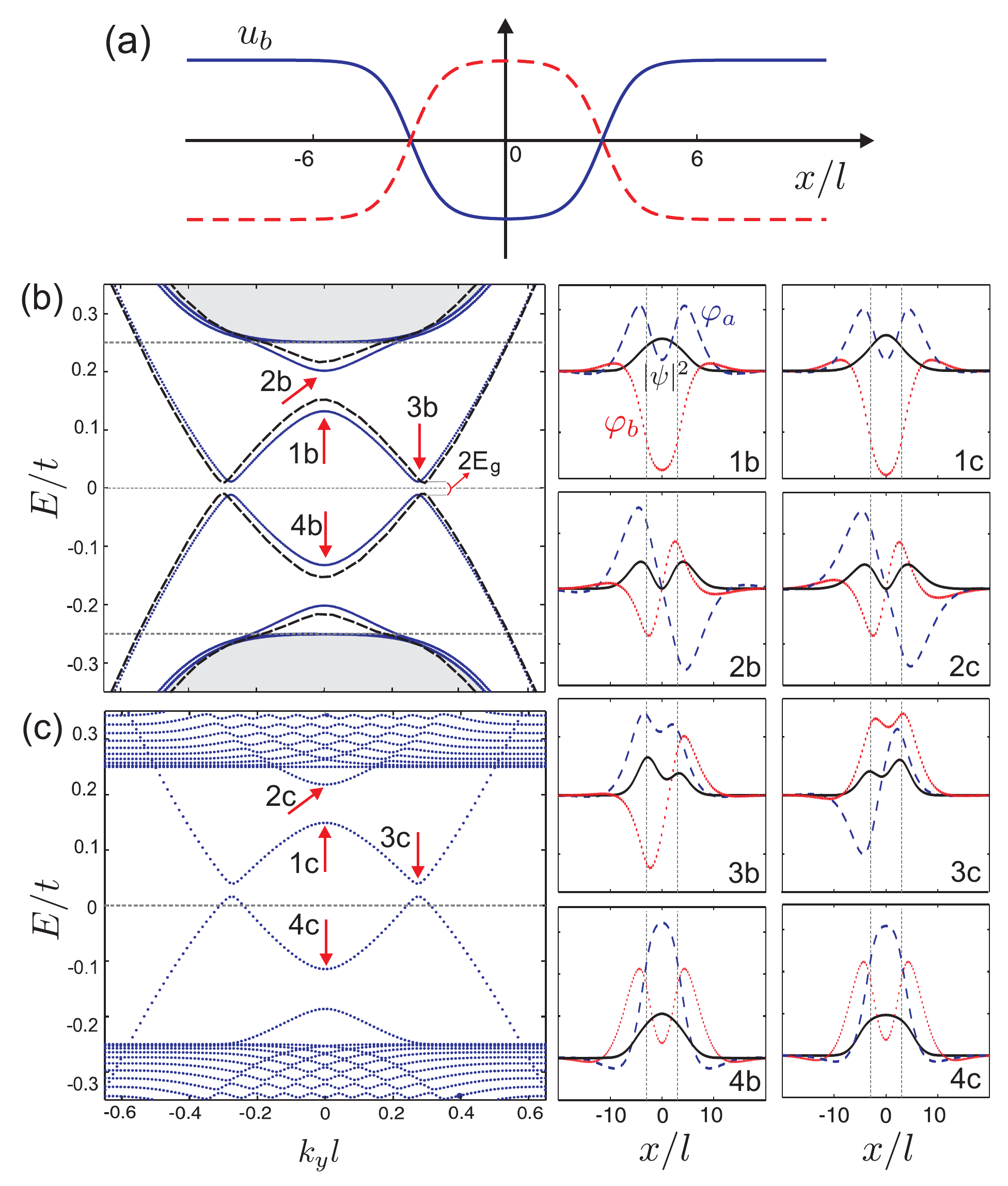}
\caption{(Color online) Energy levels of a sharp kink-antikink
profile with $u_{b}=0.25$, $d=3(\approx 5~nm)$ and $\delta=1$ (the
profile is depicted in (a)) for (b) $B=0~T$ and (c) $B=3~T$. The
black dashed curves in panel (b) display analytical results as
obtained from Eq. (11). Right panels show the real part of the
wavespinors and the corresponding probability density for the points
at the energy spectrum that are indicated by arrows in panels
(b,c).} \label{fig12}
\end{figure*}

\subsection{Magnetic field dependence}
Dependence of the energies of the 1D bound states on an external
magnetic field is shown in Fig. \ref{fig6} for (a) $k'_{y}=0$ and
(b) $k'_{y}=0.15$. In order to show the effect of a magnetic field
on the chiral states (blue solid curves) and the other localized
bound states (red dashed curves) we present the results for a smooth
potential (i.e. $\delta=10$). It is seen that the chiral states are
very weakly influenced by the magnetic field. This is a consequence
of the strong confinement of these states in the kink potential (see
Fig. \ref{Psi2S} and Fig. \ref{fig2}(a,c)). In a semiclassical view,
the movement of the carriers is constrained by the kink potential
and that together with the unidirectional propagation, prevents the
formation of cyclotron orbits. For the energy levels above the
chiral states, the energy values increase as the magnetic field
increases, because of the weaker confinement of these states as is
apparent from Figs. \ref{fig2}(b,d,e,f).
\begin{figure}
\centering
\includegraphics[width=8cm]{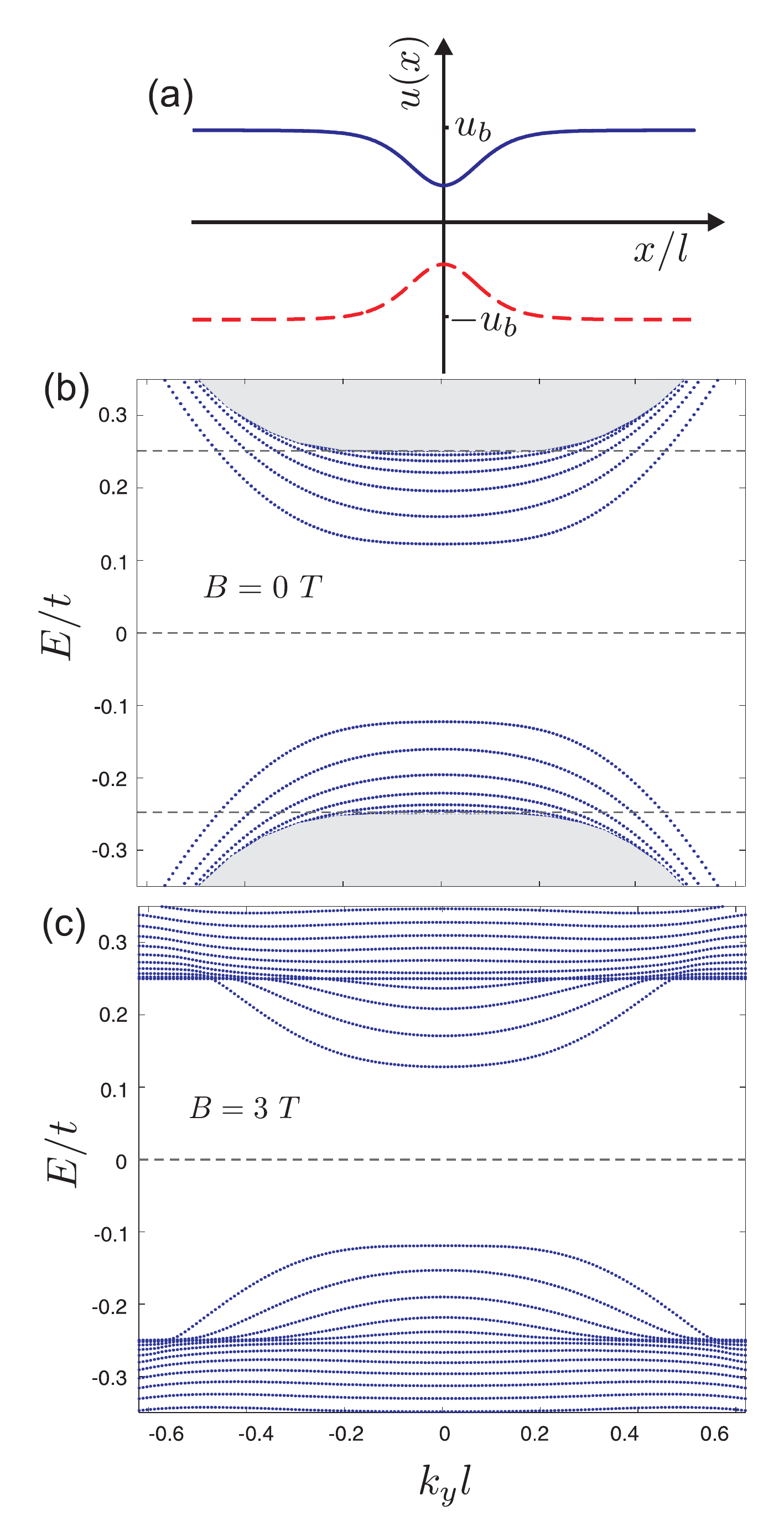}
\caption{(Color online) Energy levels of a smooth kink-antikink
profile on bilayer graphene with $u_{b}=0.25$, $d=3$ and $\delta=10$
for (a) $B=0~T$ and (b) $B=3~T$.} \label{fig14}
\end{figure}
\begin{figure}
\centering
\includegraphics[width=8.5cm]{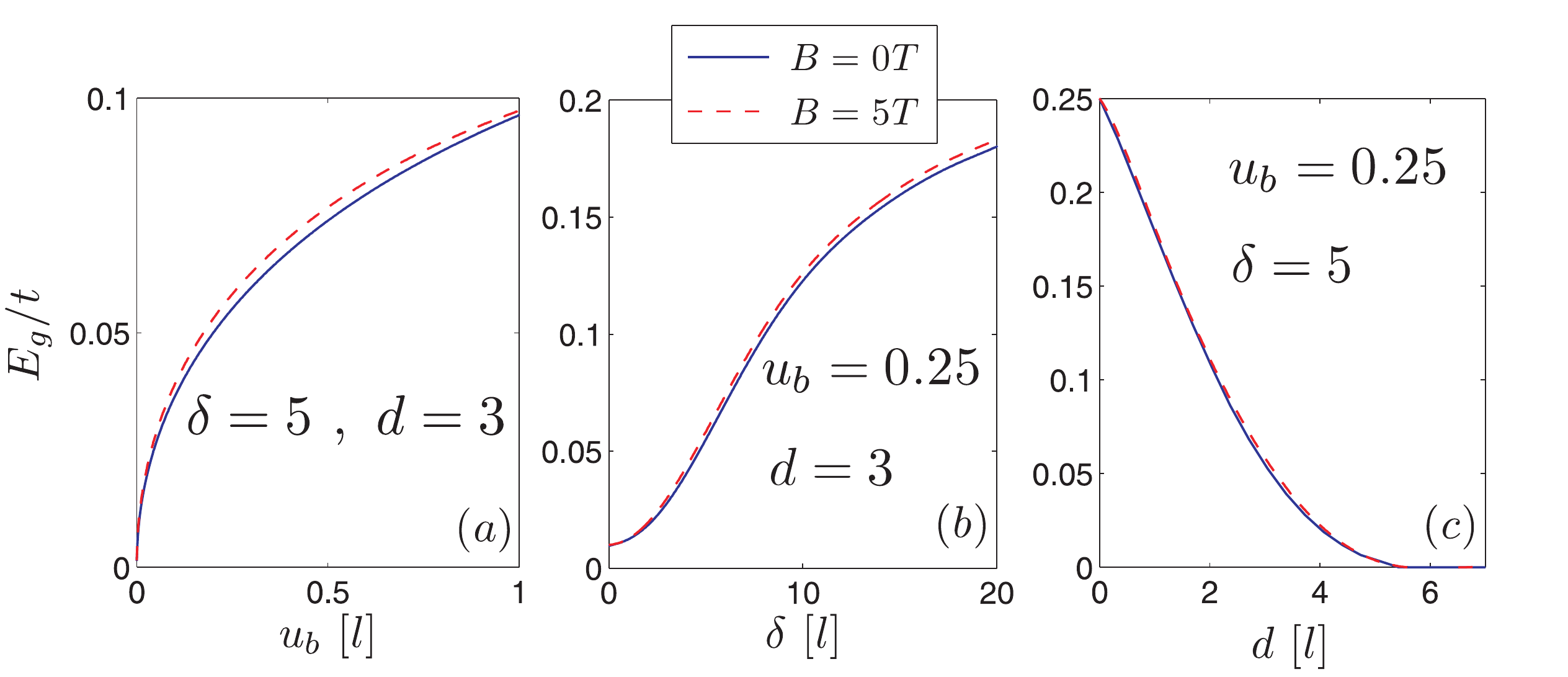}
\caption{(Color online) The energy gap $E_{g}$ (indicated in Fig.
\ref{fig12}(b)) as function of (a) maximum amplitude of the gate
voltage $u_{b}$, (b) the width of kink profile $\delta$ and (c) the
position of the kink and antikink $d$. Other parameters are shown in
the figures.} \label{fig15}
\end{figure}
\begin{figure}
\centering
\includegraphics[width=8cm]{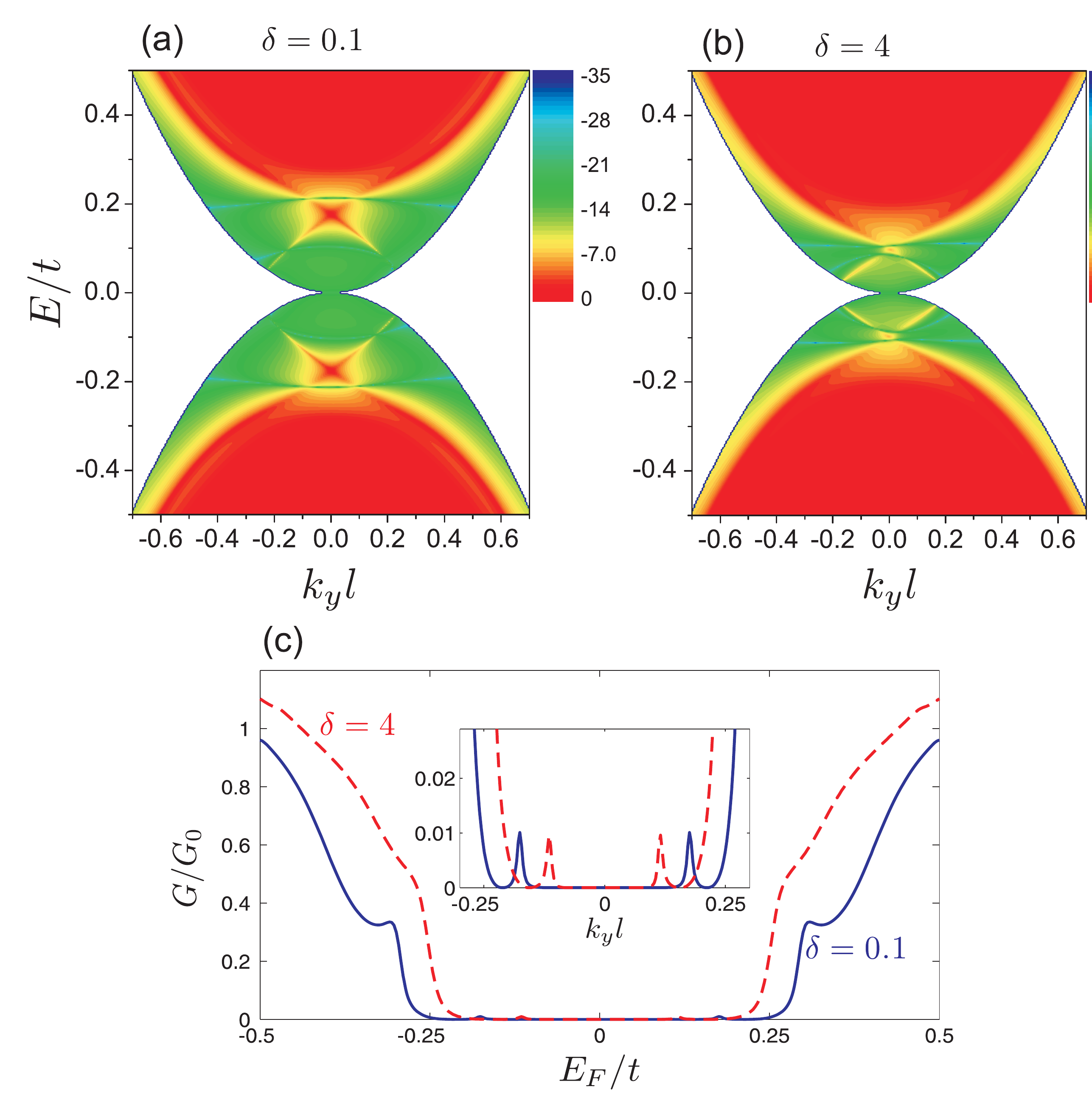}
\caption{(Color online) (a,b) Contour plot of the transmission $T$
(in logarithmic scale) through a kink-antikink structure with the
length $L=24$ ($\approx 40~nm$) for $u_{b}=0.25$ and $d=6$ with (a)
$\delta=0.1$ and (b) $\delta=4$. (c) Conductance $G/G_{0}$ vs Fermi
energy of a kink-antikink potential for the same parameters as
(a,b).}\label{CD}
\end{figure}
\begin{figure}
\centering
\includegraphics[width=8cm]{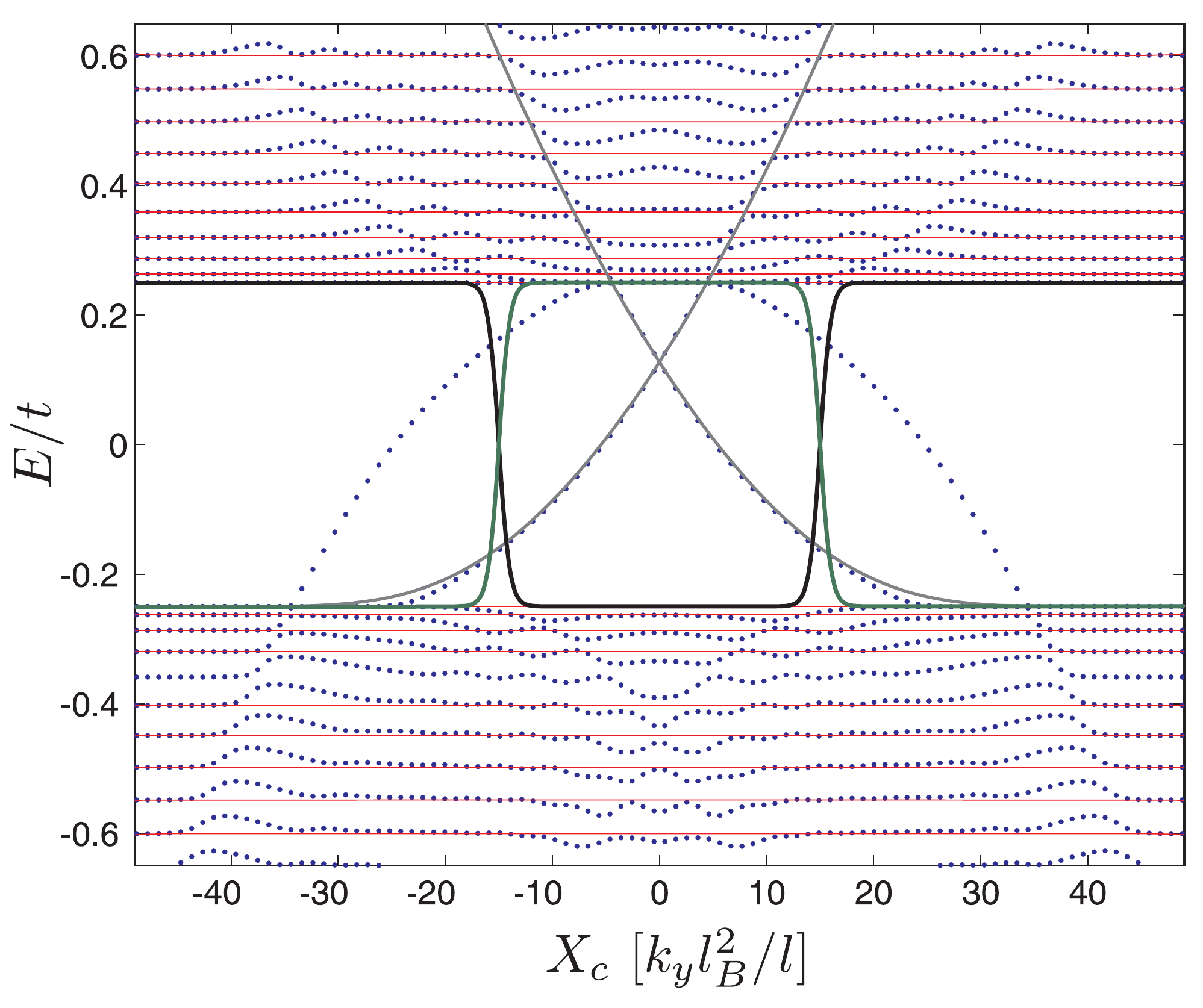}
\caption{(Color online) Energy spectrum of a kink-antikink profile
in bilayer graphene as function of the cyclotron orbit center
$X_{c}$ with $u_{b}=0.25$, $d=15$, $B=7~T$ and $\delta=1$. Solid red
lines correspond to the Landau levels of a biased BLG. The gray
solid curves show fitted functions (given in the text) to the
numerical results. The solid black and green curves describe the
potential, respectively, in the upper and lower layer.}\label{Xcd}
\end{figure}
\begin{figure}
\centering
\includegraphics[width=8cm]{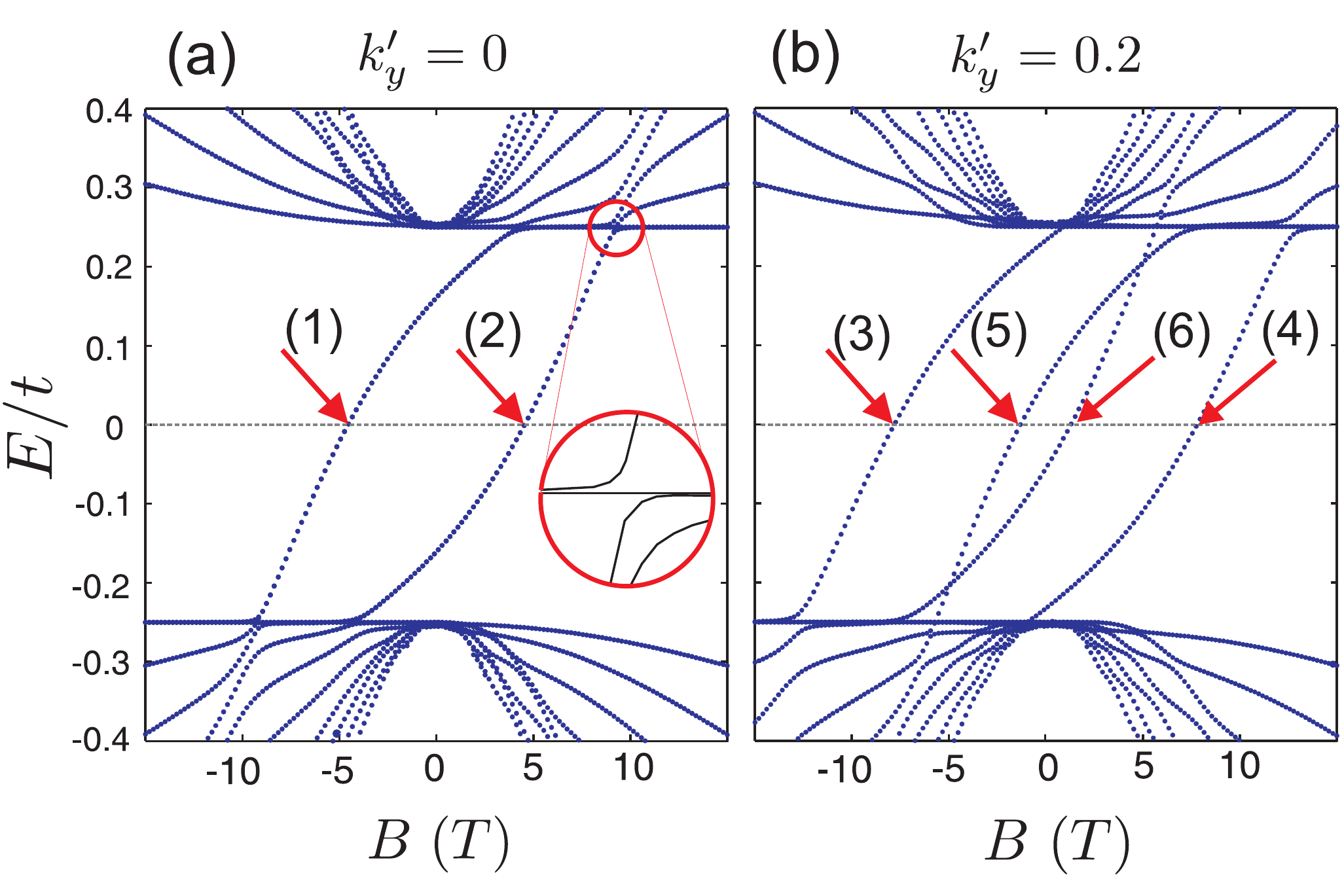}
\caption{(Color online) Energy levels of a sharp kink-antikink
profile ($\delta=1$) as function of external magnetic field for
$u_{b}=0.25$ with (a) $k'_{y}=0$ and (b) $k'_{y}=0.2$.} \label{fig9}
\end{figure}

Figure \ref{Xcs1} shows the spectrum of a sharp ($\delta=1$) single
kink potential in the presence of an external magnetic field $B=7~T$
as function of the orbit center $X_{c}=k_{y}l_{B}^{2}/l$ where
$l_{B}=\sqrt{\hbar/eB}$ is the magnetic length. The solid lines
represents the applied kink potential to upper (black) and lower
(green) layer. The results show that the topological states are
practically not affected by the magnetic field. The free energy
region (i.e. $ |\epsilon|>u_{b}$) in the absence of magnetic field
now is replaced with Landau levels (the solid red lines are the
Landau levels of a biased bilayer graphene). In some region the
Landau levels are influenced by the kink potential and
anti-crossings appear in the low energy spectrum. Some of these
anti-crossings are situated along the extension of the topological
states into the $|\epsilon|>u_{b}$ region. In addition of these
anti-crossings the Landau levels display some resonances along the
energy levels of a biased BLG (red solid curves in Fig.
\ref{fig1}(a)) which can be linked to the edge effects of the
potential profile. The position of the resonances can be fitted to
$\epsilon= a\sqrt{(X_{c}-X_{0})^{4}+(\epsilon_{n}/a)^2}$ where
$a=0.0006$ and $X_{0}=9.25$ are fitting parameters and
$\epsilon_{n}$ indicates the $n$th Landau level of a biased BLG (see
solid purple curves). Also the topological levels can be fitted to
$\epsilon= a\sqrt{(X_{c}-X_{0})^{4}+(u_{b}/a)^2}$ (dashed gray
curve) with $a=0.0003$ and $X_{0}=-26$. Panels (b,c) show the
probability densities for the points that are indicated by full red
circles in the energy spectrum. For the points on the purple solid
curves (2a,3a) the distribution of the carriers by the magnetic
field is influenced by the small confinement by the interface
potential (see solid and dashed curves in panel b). The probability
density for the point on the fitted curve along the topological
level (2b) shows a higher peak at the kink interface ($x=0$)
indicating that the kink potential acts as an attractive potential
(solid curve in panel (c)). The other probabilities are clearly
those of free electron LL. The result for a smooth kink potential
$\delta=10$ is shown in Fig. \ref{Xcs10} where the energy value of
the extra bound states are increased by the magnetic field and the
topological levels are practically not affected by the magnetic
field. 
\begin{figure}
\centering
\includegraphics[width=8cm]{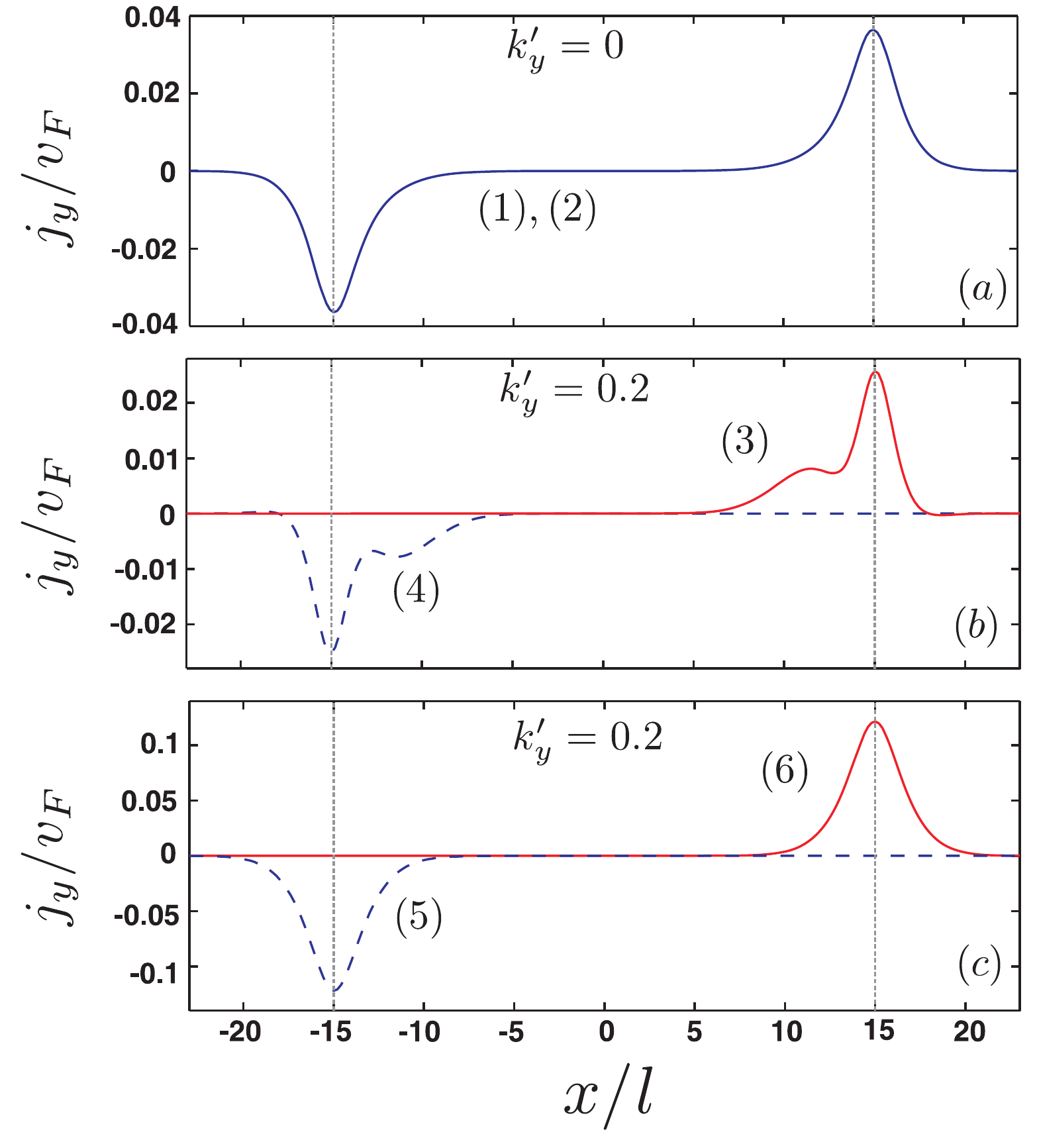}
\caption{(Color online) $y$-component of the persistent current of a
kink-antikink profile in bilayer graphene as function of the
$x$-direction for $E=E_{F}$ and for values of the magnetic field
that are indicated by $(1),(2),...$ in Figs. \ref{fig9}(a,b).}
\label{fig10}
\end{figure}

\begin{figure}
\centering
\includegraphics[width=8cm]{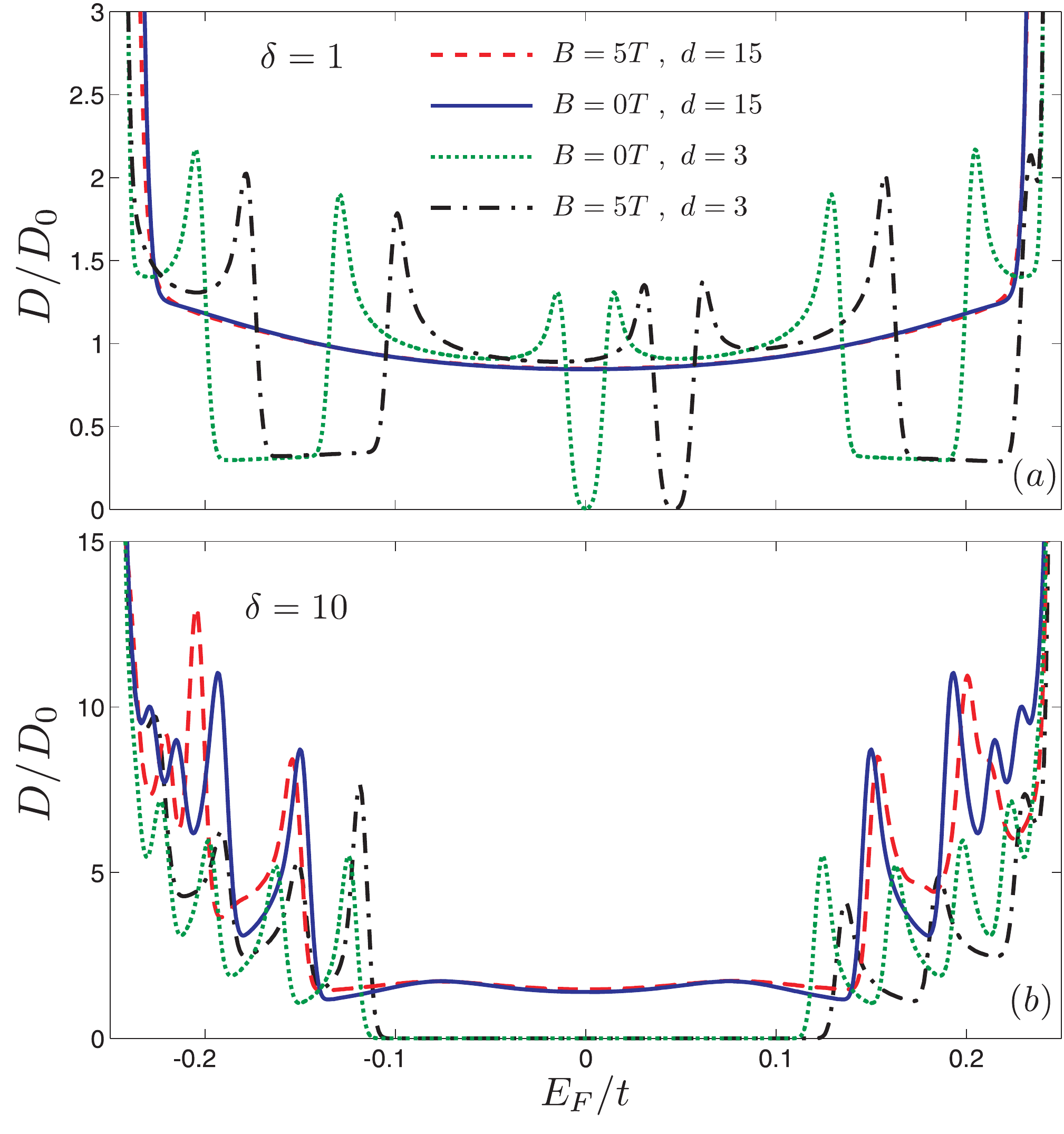}
\caption{(Color online) Density of states (DOS) of kink-antikink
potentials for (a) $\delta=1$ and (b) $\delta=10$ with $u_{b}=0.25$.
The results are presented for $d=3$ and $d=15$ with $B=0~T$ and
$B=5~T$.}\label{DOSd}
\end{figure}

The localization of the states is reflected in the position dependence of the current.
The current in the $y$-direction is obtained using
\begin{equation}
j_{y}=iv_{F}[\Psi^{\dag}(\partial_{x}\sigma_{y}-\partial_{y}\sigma_{x})\Psi
+\Psi^{T}(\partial_{x}\sigma_{y}+\partial_{y}\sigma_{x})\Psi^{\ast}]
\end{equation}
where $\Psi = [\psi_a(x,y)~,~\psi_b(x,y)]^T$. By substituting $\Psi(x,y)=e^{ik_{y}y}[\varphi_{a}(x),\varphi_{b}(x)]^{T}$ we have
\begin{equation}
j_{y}=2v_{F}\big[Re\{\varphi_{a}^{\ast}\partial_{x}\varphi_{b}-\varphi_{b}^{\ast}\partial_{x}\varphi_{a}\}
+2k_{y}Re\{\varphi_{a}^{\ast}\varphi_{b}\}\big].
\end{equation}
The $x$-component of the current vanishes for the confined states.
In Fig. \ref{jyS}, the $y$-component of the persistent current for a
sharp (blue curves) and smooth (black curves) single potential kink
profile is shown as function of the $x$ direction without magnetic
field (solid curves) and in the presence of the magnetic field
(dashed curves). In the absence of a magnetic field the current is
localized around $x=0$ for both sharp ($\delta=1$) and smooth
$(\delta=10)$ potentials. For a smooth profile the wavefunction of
the topological states and consequently also the current density
profile is broadened (compare Figs. \ref{fig1}(b,c) with Figs.
\ref{fig2}(a,c)). A magnetic field shifts the density profile
slightly to the right (see the inset of Fig. \ref{jyS}) due to the
Lorentz force and there is also a very small narrowing of the
current distribution.

Next we consider the density of states (DOS) for the kink potential. The number of k-states per unit energy is given by
\begin{equation}
\displaystyle{D(E)=\frac{D_{0}}{2\pi}\sum_{n}\int}dk_{y'}\delta(\epsilon-\epsilon_{n,k'_{y}}).
\end{equation}
where $D_{0}=(\hbar v_{F})^{-1}$. To calculate the DOS numerically
we introduce a Gaussian broadening,
\begin{equation}\label{delta}
\displaystyle{\delta(\epsilon-\epsilon_{n,k'_{y}})\rightarrow \frac{1}{\Gamma\sqrt{\pi}} \exp[-\frac{(\epsilon-\epsilon_{n,k'_{y}})^2}{\Gamma^2}],}
\end{equation}
where $\Gamma$ is the broadening which is taken as $\Gamma=0.005$ in
our calculations. Figure \ref{DOSs} shows the DOS as function of
Fermi energy $E_{F}$ in the absence and presence of an external
magnetic field for sharp ($\delta=1$) and smooth ($\delta=10$) kink
potentials. For a sharp profile the topological levels contribute an
almost constant value to the DOS for $|\epsilon|<u_{b}$ even in the
presence of an external magnetic field. For the smooth profile,
peaks corresponding to the non-topological levels appear in the DOS
and note that only these peaks are shifted in the presence of a
magnetic field while the DOS of the topological states are not
affected by the magnetic field.

We now turn to the transport properties of a kink potential and look
at the influence of the topological states on the conductivity in
the $y$-direction $(\sigma_{yy})$. For elastic scattering the
diffusive conductivity $\sigma_{yy}$ is given by\cite{charbonneau},
\begin{equation}
\displaystyle{\sigma_{yy}=\frac{e^2 v_{F}}{2\pi\hbar k_{B}T}\sum_{n}\int dk'_{y}\tau v_{n,y}^{2}f_{n,k'_{y}}(1-f_{n,k'_{y}})}.
\end{equation}
Here $T$ is the temperature, $v_{n,y}=\partial\epsilon_{n}/\partial
k'_{y}$ is the electron velocity, $\displaystyle{
f_{n,k}=1/[\exp\big((E_{F}-E_{n,k'_{y}}\big)/K_{B}T)+1]}$ is the
equilibrium Fermi-Dirac distribution function, and $\tau$ is the
momentum relaxation time. For low temperatures we assume that $\tau$
is approximately constant, evaluated at the Fermi level
($\tau\approx\tau_{F}$), and replace the product
$f_{n,k'_{y}}(1-f_{n,k'_{y}})/k_{B}T$ by the delta function given in
Eq. (\ref{delta}). The results are presented as function of $E_{F}$
in Fig. \ref{sigmaS} in the units of
$\sigma_{0}=e^{2}v_{F}\tau_{F}/\hbar L_{y}$ for both sharp
($\delta=1$) and smooth ($\delta=10$) potentials with $B=0~T$ and
$B=5~T$. Due to the robust confinement of the topological levels the
conductivity is constant in the energy gap even for a non-zero
magnetic field (solid blue curve for $B=0~T$ and red dashed curve
for $B=5~T$). The extra localized levels in the case of $\delta=10$
lead to an increasing conductivity as function of $E_{F}$. Note that
in the presence of an external magnetic field some of the additional
electron(hole) states are shifted up(down) in energy (see Figs.
\ref{fig2} and \ref{Xcs10}) which results in smaller $\sigma_{yy}$
at the $|\epsilon_{F}|<u_{b}$ region compared to the conductivity in
the absence of magnetic field (black dotted-dashed curve).

\section{Kink-antikink}
\subsection{Zero magnetic field}

Next we consider a potential profile with a pair kink-antikink. The kink-antikink potential is modeled by,
\begin{equation}\label{eq}
\displaystyle{u(x')=u_{b}\big[\tanh(\frac{x'-d}{\delta})-\tanh(\frac{x'+d}{\delta})+1\big]}
\end{equation}
where, $d$ is the distance between the kink and the antikink in
units of $l$. The spectrum of the localized states in the absence of
a magnetic field is shown in Fig. \ref{fig8}(a) for $u_{b}=0.25$,
$\delta=1$ and $d=15$. The black dashed curves are the analytical
results for $\delta=0$ which are obtained using Eqs. (\ref{eq88}).
Note that there are only two chiral states per kink which leads to
the appearance of crossing points in the energy spectrum (at $E=0$
and $k'_{y}=0$). The spinor components and probability densities
associated with the points indicated inside the circle in Figs.
\ref{fig8}(a) are shown in the panels (1a, 2a,...,5a). In the
absence of a magnetic field and for the points around the energy
level crossing the carriers are strongly confined at either the
position of the kink or antikink. The wavefunction corresponding to
an energy at the crossing point (panel 5a) is localized at both the
kink and antikink.

Next we investigate smooth potential kink  profiles. In Figure
\ref{fig11}(a) the energy spectrum of a smooth kink-antikink profile
(i.e. $\delta=10$) is presented for zero magnetic field. As in the
case of the single kink profile additional bound states appear in
the energy spectrum. The overlap between theses states leads to the
appearance of crossing points in the energy spectrum. The
wavespinors and the corresponding probability density for the points
indicated by arrows in panel (a) are shown in panels (1a,2a,3a,4a).
In the absence of a magnetic field and for $k'_{y}=0$ the states are
localized at both kink and antikink (panels 1a, 2a and 4a) whereas,
panel 3a shows that the confinement tends to the kink or antikink at
$k'_{y}\neq0$.

Decreasing the distance between the kink and antikink generate an
unperfect kink-antikink profile\cite{unperfect}. This profile is
illustrated in Fig. \ref{fig12}(a). The energy spectrum of such a
profile is shown in Fig. \ref{fig12}(b) for $B=0~T$, $\delta=1$ and
$d=3(\approx 5~nm)$. The analytical results (obtained from Eq.
(\ref{eq88})) for $\delta=0$ are shown by the black dashed curves.
Now the crossing points in the energy spectrum for the case of
$d=15$ (see Fig. \ref{fig8}(a)) are replaced with anticrossings and
an energy gap $E_{g}$ appears in the energy spectrum. The positions
of these minigaps move when we increase the magnetic field as in
apparent from Fig. \ref{fig12}(c). The panels (1b,2b,3b,4b) show the
real parts of the wavespinors and corresponding probability density
for the indicated points in Fig. \ref{fig12}(b) by red arrows. Note
that due to the decreasing distance of the kink and antikink the
carriers can be localized between the kink and antikink.

Figure \ref{fig14}(a) displays the energy spectrum of a smooth
($\delta=10$) kink-antikink potential with $d=3$ for $B=0~T$. Now
the kink and antikink are close to each other and the smoothness of
the potential leads to extra localized levels. Therefore the
crossing and anticrossing points between the additional bound states
are seen to disappear and the energy gap between the topological
levels is increased. The magnitude of the energy gap $E_{g}$ depends
on the width of the interface region $\delta$, the maximum value of
the potential $u_{b}$ and the distance between the kink and
antikink. This is shown in Fig. \ref{fig15}, where $E_{g}$ is
plotted as function of $u_{b}$, $\delta$ and $d$ respectively in
panels (a), (b) and (c) in the absence of magnetic field (blue solid
curves). As shown in panels (a,b) the energy gap is an increasing
function of $u_{b}$ and $\delta$. When $\delta$ increases the first
energy level at the spectrum changes from a Mexican hat shape to a
parabola. Therefore, $E_{g}$ increases with increasing $\delta$
(compare the potentials illustrated in Figs. \ref{fig12}(a) and
\ref{fig14}(a)). Increasing the distance of the kink and antikink
results in perfect unidirectional states and the gap disappears
(panel \ref{fig15}(c)).

Next we consider the transmittance of a kink-antikink potential. In
Fig. \ref{CD} we show a contour plot of the transmission probability
(in logarithmic scale) for the kink-antikink structure with
$L_{x}=24$ for (a) $\delta=0.1$ (sharp) and (b) $\delta=4$ (smooth)
potentials. The results show a nonzero region for the transmittance
below the gap where the topological levels corresponding to the kink
and antikink cross each other. The conductance as function of Fermi
energy is plotted in Fig. \ref{CD}(b). A small region of
transmittance appears in the energy gap due to the chiral states
that appears as small peaks in the conductance (see the inset of
panel (c)).

\subsection{Magnetic field dependence}
Figure \ref{fig8}(b) shows the kink-antikink energy levels in the
presence of an external magnetic field ($B=3~T$). The results show a
shift of the four intra-gap energy branches as the magnetic field
increases. In addition, the continuum of free states at zero
magnetic field (shadowed region in Fig. \ref{fig8}(a) is replaced by
a set of Landau levels. The spinor components and probability
densities associated with the points indicated inside the circle in
Figs. \ref{fig8}(b) are shown in the panels (1b,2b,...5b). For
non-zero magnetic field the states show a shift of the probability
density towards the region between the kink and the antikink. This
is caused by the additional confinement due to the magnetic field.

The energy levels of a smooth kink-antikink profile (i.e.
$\delta=10$) in the presence of a perpendicular magnetic field is
presented in Fig. \ref{fig11}(b). Now the crossings points (in the
case of $B=0~T$) changed into anticrossings. In the inset of Fig.
\ref{fig11}(b) an anti-crossing is enlarged. Due to the strong
confinement of the potential the magnetic field can only lead to a
shift up in energy of the localized chiral states. The wavespinors
and the corresponding probability density for the points indicated
by arrows in panel (b) are shown in the panels (1b,2b,3b,4b). In the
presence of an external magnetic field and at the crossing points of
the topological states (panels 1b, 2b) due to the strong confinement
by the potential the magnetic field can only affect weakly the
electrons. At the first anticrossing (panels 3b and 4b) which arises
form the overlap of the first bound states in the kink and antikink
potentials, the electrons are confined closer to the center of the
potential. 

The energy levels for a sharp ($\delta=1$) kink- antikink potential
with $d=3$ is presented in Fig. \ref{fig12}(c). The crossings which
appeared in the energy spectrum due to the overlap of the extra
bound states in the absence of magnetic field (see Fig.
\ref{fig12}(b)) now are replaced with anti-crossings and  the energy
gap $E_{g}$ between the kink and antikink states is shifted up in
energy due to the confinement by the magnetic field. Panels
(1c,2c,3c,4c) show the wavespinors and probability density for the
points indicated by arrows in panel (c). The energy spectrum of a
smooth kink-antikink potential with $d=3$, $\delta=10$ and in the
presence of an external magnetic field $B=3~T$ is shown in Fig.
\ref{fig14}(b). In the presence of a magnetic field the energy gap
is shifted and the symmetry of the spectrum around $E=0$ for $B=0~T$
is broken (see panel (a)). The energy gap in the presence of a
magnetic field ($B=5~T$) is shown in Fig. \ref{fig15} as red dashed
curves. Notice that an external magnetic field only shifts up the
energy gap and the gap size remains constant.

The energy spectrum of a kink-antikink potential is shown in Fig.
\ref{Xcd} as function of orbit center $X_{c}$ for $\delta=1$,
$u_{b}=0.25$ and $d=15$. The kink-antikink potential is depicted in
the figure by the dashed curves. Such as for the single kink
potential the topological levels can be fitted to
$\displaystyle{\epsilon_{\pm}\approx a\sqrt{(X_{c}\pm
X_{0})^4+(u_{b}/a)^2}}$ (see gray solid curves) where $-(+)$
corresponds to the kink(antikink) branches ($a=0.003$ and $X_{0}=41$
are the fitting parameters). Now the Landau levels above the gap are
affected by the kink-antikink potential where anti-crossing points
appear along the topological levels. The solid red lines are the
Landau levels in a biased BLG.

Figure \ref{fig9} shows the dependence of the energies on the
external magnetic field for (a) $k'_{y}=0$ and (b) $k'_y = 0.2$. The
branches that appear for $\displaystyle{|\epsilon|>0.25}$ correspond
to Landau levels that arise from the continuum of free states. For
the kink-antikink case, however, the overlap between the states
associated with each confinement region allows the formation of
Landau orbits. Therefore, the proximity of an antikink induces a
strong dependence of the states on the external field.

Figure \ref{fig10} shows plots of the $y$-component of the current
density as function of $x$ for the states labeled (1) to (6) in
panels (a) and (b) of Fig. \ref{fig9}. It should be noticed that a
non-zero current can be found for $E = 0$ and $k'_{y}\neq 0$, as can
be deduced from the dispersion relations. For $k'_y = 0$ the results
presented in Fig. \ref{fig10}(a) show a persistent current carried
by electrons localized at each kink region, irrespective of the
direction of $\boldsymbol{B}$, as exemplified by the states (1) and
(2) which correspond to opposite directions of the magnetic field.
For non-zero wave vector, however, as shown in panels (b) and (c),
the current is strongly localized around one of the potential kinks.
In Fig. \ref{fig10}(b), the current density curve shows an
additional smaller peak caused by the strong magnetic field ($B
\approx 7.5~T$) where, the carriers can also be confined closer to
the center.

The density of states of the topological states
for (a) $\delta=1$ and (b) $\delta=10$ kink-antikink potential is shown in
Fig. \ref{DOSd} with $d=3$ and $d=15$. The results show additional peaks for a
sharp kink-antikink with $d=3$ which is due to the splitting of the topological levels. Note that
the energy gap leads to a zero density at $E_{F}=0$ for zero magnetic field (blue circles in (a)) while shifting
the gap in the presence of a magnetic field results in a
non-zero DOS at $E_{F}=0$ (red diamonds in (a)). For the
smooth profiles the non-topological 1D states lead to the
appearance of additional peaks in the DOS (panel (b)) that shift with the magnetic field.\\

\section{Concluding remarks}
In summary we obtained the energy spectrum, the density of states, the
transmission and conductivity for carriers moving in BLG in the presence of
asymmetric potentials (i.e. kink and kink-antikink profiles) in each layer of
the BLG. Uni-directional chiral states are localized at the location of the
kink (or antikink). By controlling the gate voltages and/or the smoothness
of the kink profile the number of one-dimensional metallic channels and
their subsequent magnetic response can be configured.\\
\indent The effect of an external magnetic field perpendicular to
the bilayer sheet was investigated. We found that the influence of
the magnetic field is very different for single and double kinks. Due to
the strong confinement by the kink potential, the topological states are
weakly affected by the magnetic field in the case of a single kink profile.\\
\indent Changing the sign of the kink potential smoothly (i.e. broadening the kink potential) leads
to extra bound states which have a very different behavior as compared to the uni-directional
topological states. First, these states are no longer uni-directional and they have a quasi-1D
free electron-type of spectrum which is asymmetric around $k_{y}=0$. Second, they are less
strongly localized at the kink of the potential as compared to the chiral states and their
probability distribution appears as those of excited states of the chiral state.\\
\indent In the case of parallel kink-antikink profiles apparent crossings of
the energy levels are found in the spectrum. Decreasing the distance between (and/or smoothing)
the kink-antikink profiles turns into anti-crossings. It opens a gap in the topological
state spectrum. This allows for a robust 1D system having a tunable minigap.

\section{Acknowledgment}
This work was supported by the Flemish Science Foundation (FWO-Vl),
the Belgian Science Policy (IAP), the European Science Foundation
(ESF) under the EUROCORES program EuroGRAPHENE (project CONGRAN),
the Brazilian agency CNPq (Pronex), and the bilateral projects
between Flanders and Brazil and the collaboration project FWO-CNPq.

\end{document}